\documentclass[usenatbib]{mn2e}
\usepackage{times}

\pdfoutput=1

\usepackage{natbib}
\usepackage[T1]{fontenc}
\usepackage{amsmath}
\usepackage{amssymb}
\usepackage{bm}
\usepackage{booktabs}
\usepackage{graphicx}
\usepackage{mathtools}
\usepackage{color}

\newcommand{\refig}[1]{Fig.~\ref{#1}}
\newcommand{\reftab}[1]{Table~\ref{#1}}
\newcommand{\tx}[1]{\textrm{#1}}

\newcommand{\modela}{H256r}		
\newcommand{\modelb}{Ml256r}		
\newcommand{\modelc}{Mm256r}		
\newcommand{\modeld}{Mh256r}		
\newcommand{\modelB}{Ml512r}		
\newcommand{\modelC}{Mm512r}		
\newcommand{\modelD}{Mh512r}		
\newcommand{\modele}{Mh256r-pi}	
\newcommand{\modelA}{H512m1}		
\newcommand{\modelBm}{Ml512m1}	

\newcommand*{\ks}[1]{#1_{\rm \mbox{\tiny KS}}}
\newcommand*{\bl}[1]{#1_{\rm \mbox{\tiny BL}}}
\def\be{\begin{equation}}
\def\ee{\end{equation}}

\title[PPI suppression by MRI in relativistic discs]{Papaloizou-Pringle instability suppression by the magnetorotational instability in relativistic accretion discs}
\author[M. Bugli et al.]
{M.~Bugli$^{1,2}$\thanks{matteo@mpa-garching.mpg.de}, J.~Guilet$^{1,3,4}$, E.~M\"uller$^{1,2}$, L.~Del~Zanna$^{5,6,7}$, N.~Bucciantini$^{6,5,7}$, P.~J.~Montero$^{1}$ \\
$^1$ Max-Planck-Institute f\"ur Astrophysik, Karl-Schwarzschild Strasse 1, 85741 Garching, Germany \\
$^2$ Technische Universit\"at M\"unchen, Physik Department \\
$^3$ Max-Planck Princeton Center of Plasma Physics\\
$^4$ Laboratoire AIM, CEA/DRF-CNRS-Universit{\'e} Paris Diderot, IRFU/D{\'e}partement d'Astrophysique, CEA-Saclay, F-91191, France \\
$^5$ Dipartimento di Fisica e Astronomia, Universit\`a di Firenze, Via G. Sansone 1, 50019 Sesto F.no (Firenze), Italy \\
$^6$ INAF -- Osservatorio Astrofisico di Arcetri, L.go E. Fermi 5, 50125 Firenze, Italy\\
$^7$ INFN -- Sezione di Firenze, Via G. Sansone 1, 50019 Sesto F.no (Firenze), Italy }
\date{}

\begin{document}

\maketitle
\begin{abstract}
Geometrically thick tori with constant specific angular momentum have been widely used in the last decades to construct numerical models of accretion flows onto black holes. Such discs are prone to a global non-axisymmetric hydrodynamic instability, known as Papaloizou-Pringle instability (PPI), which can redistribute angular momentum and also lead to an emission of gravitational waves. It is, however, not clear yet how the development of the PPI is affected by the presence of a magnetic field and by the concurrent development of the magnetorotational instability (MRI). We present a numerical analysis using three-dimensional GRMHD simulations of the interplay between the PPI and the MRI considering, for the first time, an analytical magnetized equilibrium solution as initial condition. In the purely hydrodynamic case, the PPI selects as expected the large-scale $m=1$ azimuthal mode as the fastest growing and non-linearly dominant mode. However, when the torus is threaded by a weak toroidal magnetic field, the development of the MRI leads to the suppression of large-scale modes and redistributes power across smaller scales. If the system starts with a significantly excited $m=1$ mode, the PPI can be dominant in a transient phase, before being ultimately quenched by the MRI. Such dynamics may well be important in compact star mergers and tidal disruption events.
\end{abstract}

\section{Introduction}
Accretion onto black holes provides one of the most efficient mechanisms to power up high-energy astrophysical sources such as active galactic nuclei \citep{Rees:1984,Marconi:2004,Reynolds:2014}, X-ray binaries \citep{Narayan:1995,Fender:2004,Remillard:2006} and gamma-ray bursts \citep{Woosley:1993,Piran:1999,Kumar:2015}, just to cite a few. 
The conservation of angular momentum commonly leads to the formation of accretion discs, which in order to convert gravitational binding energy into thermal, kinetic or magnetic energy need to lose their angular momentum sufficiently fast. 
 
Since black hole-disc models were proposed as central engine for quasars by \citet{Lynden-Bell:1969}, there has been a continuous interest in the physics underlying accretion onto black holes (for a general review, see \citet{Abramowicz:2013}). The seminal papers by \citet{Shakura:1973} and \citet{Lynden-Bell:1974} first described what now is known as the \emph{standard disc model}: a geometrically thin, optically thick Keplerian disc where the accretion process is driven by a local turbulent viscosity that is parametrized by the quantity $\alpha$. Given the great success of the standard disc model in providing a self-consistent way to enable accretion along with accurate predictions for the observed emission, the actual nature and physical mechanism behind the parameter $\alpha$ (which essentially gives an estimate of the efficiency of the angular momentum transport in the disc) has been the object of numerous studies and is currently still under investigation. The capability of hydrodynamic Keplerian disc models to explain from first principles the onset of a turbulent accretion flow has been debated for decades. They are indeed stable to local linear perturbations, since their distribution of specific angular momentum increases with distance from the central object (this is the well-known Rayleigh stability criterion), and in general, a small displacement of a fluid element will lead to epicyclic oscillations \citep{Kato:2016}. The non-linear stability of hydrodynamic discs is to this date a matter of debate. The first series of shearing box simulations tackling this problem \citep{Balbus:1996,Hawley:1999} suggested that hydrodynamic accretion flows do not present a transition to a turbulent state, but it has been argued by \citet{Longaretti:2002} that their results were due to a lack of resolution. More recently \citet{Lesur:2005}  showed that at high enough Reynolds numbers non-linear perturbations could lead to self-sustained turbulence, which, however, would be too weak to explain observed accretion rates. 

\begin{figure}
 \includegraphics[scale=0.3]{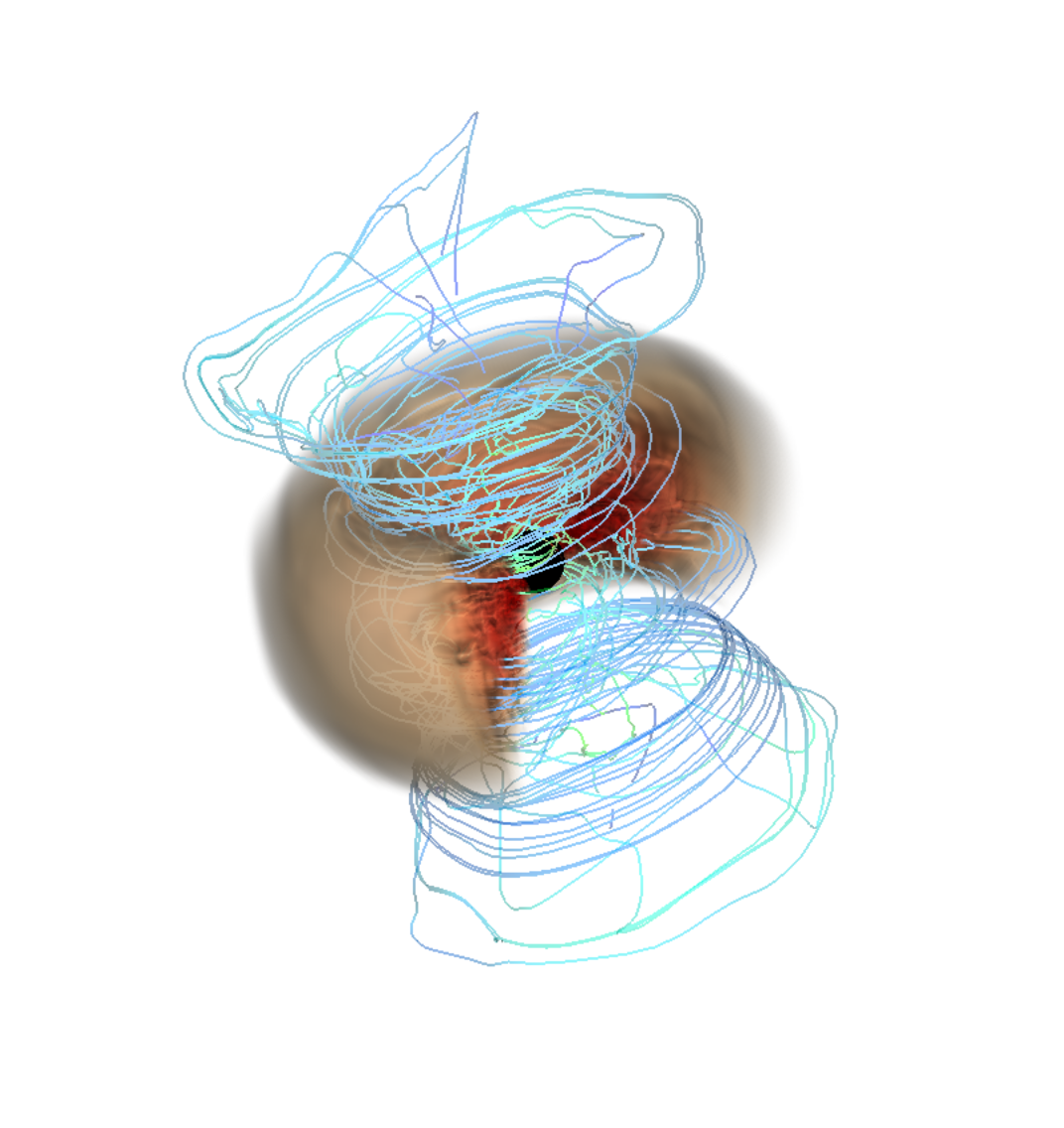}
 \caption{Volume rendering of the rest mass density of model \modelb. The stream lines represent the fluid velocity colour-coded according to the strength of the magnetic field. \label{fig:rendering}}
\end{figure}

\begin{figure}\centering
    \includegraphics[scale=0.10]{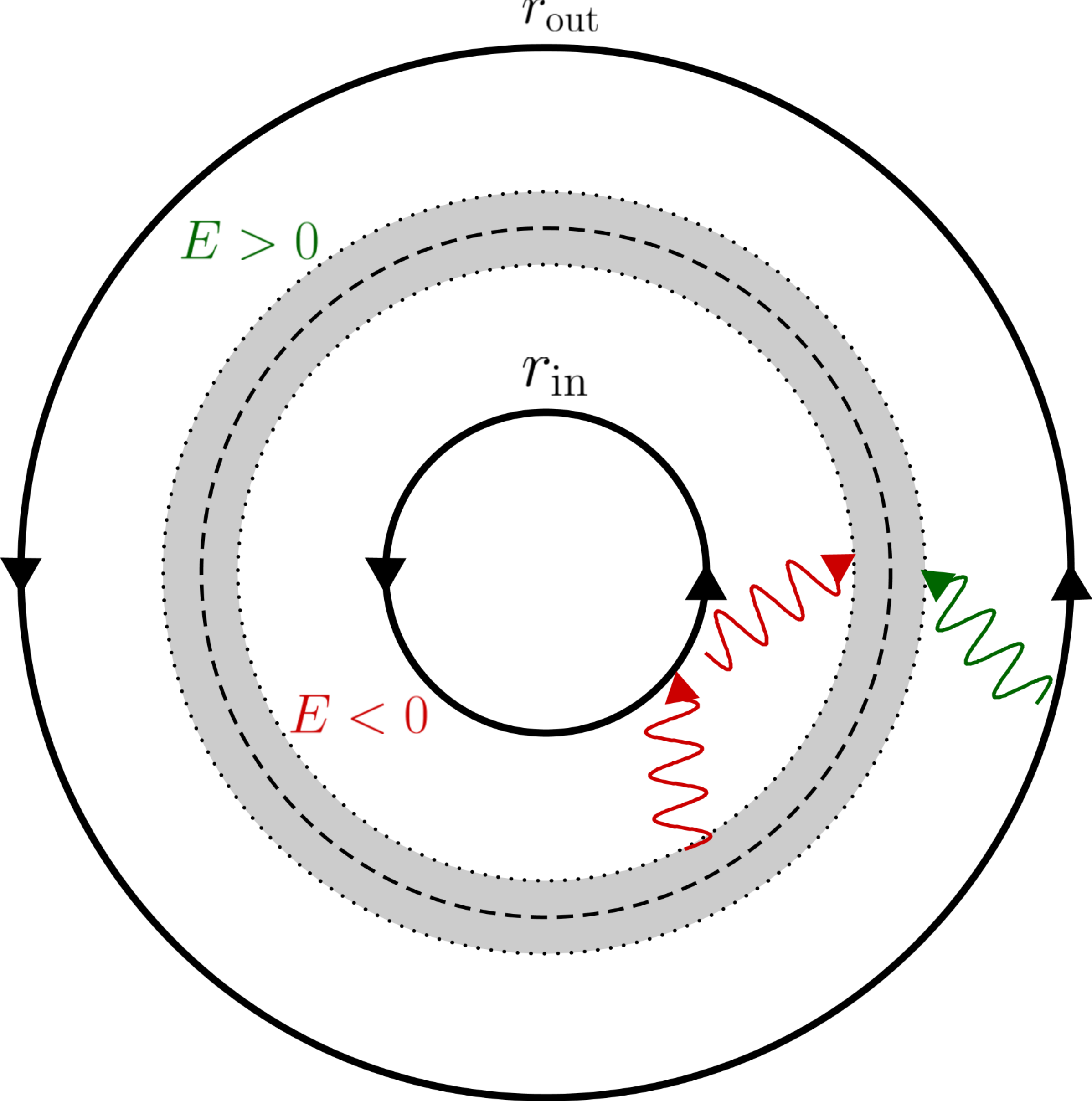}
    \caption{Schematic representation of the mechanism responsible for the PPI unstable modes. The negative-energy waves from the inner part of the disc (where $\omega<\Omega$) interact with the positive-energy ones from the outer disc (where $\omega>\Omega$) redistributing angular momentum across the corotation radius with a positive feedback by the reflective boundaries. The gray area represents the \emph{forbidden zone} surrounding the corotation radius (dashed circle), where the waves become evanescent.}\label{fig:ppi_sketch}
\end{figure}

An important class of hydrodynamic discs often used in the context of accretion onto black holes is constituted by accretion tori \citep{Abramowicz:1978}, also referred to as \emph{Polish doughnuts} (\refig{fig:rendering}). These thick discs have a large internal energy, and they rely on pressure gradients to support the disc together with centrifugal forces, resulting in a significant vertical thickening of the disc and a departure from a Keplerian distribution of specific angular momentum. Despite their local stability, \cite{Papaloizou:1984} discovered that they are prone to develop a global non-axisymmetric instability (known as Papaloizou-Pringle instability, PPI, see \refig{fig:ppi_sketch}) which is able to transport angular momentum outwards. Although capable of triggering some accretion, the PPI can not explain in a satisfactory way the ubiquity of accreting systems, since it mainly affects nearly constant angular momentum tori \citep{Goldreich:1986,Blaes:1988}. The PPI does also not fit well in the standard disc model (which assumes locally generated turbulence to enable accretion), as the instability tends to saturate in strong spiral pressure waves in radially wide, nearly-constant-angular-momentum tori \citep{Hawley:1991,De-Villiers:2002}.

The breakthrough in accretion theory was the realization that magnetic fields are the key to explain how discs can get rid of their angular momentum. The discovery of the \emph{magnetorotational instability} (MRI) in astrophysics by \cite{Balbus:1991} provided a local mechanism, efficient for a wide range of magnetic field strength, which leads to a growth of linear perturbations on dynamical time-scales and naturally develops MHD turbulence. Since then, the properties of the MRI have been studied in great detail, from both a local and a global point of view \citep{Balbus:1998,Fromang:2013,Blaes:2014}. 

Despite the fundamental importance of magnetic fields  in providing a general and universal mechanism to enable accretion in astrophysical discs, the PPI is still quite relevant as an agent of global non-axisymmetric instability, since thick discs with sub-Keplerian angular momentum distributions are expected to form in binary neutron stars \citep{Rezzolla:2010,Kiuchi:2010} or black hole-neutron star \citep{Shibata:2006,Foucart:2012} mergers, after the rotational gravitational collapse of massive stars \citep{MacFadyen:1999,Aloy:1999} and in tidal disruption events \citep{Loeb:1997,Coughlin:2014}. The stability of such wide tori has been studied from both an analytical \citep{Goldreich:1986,Glatzel:1987} and numerical \citep{Blaes:1988,Hawley:1991,De-Villiers:2002} point of view, and they have been proven to be quite generally unstable to some non-axisymmetric mode induced by the PPI. Moreover, a residual kick velocity of the central black hole after the merger can excite large-scale spiral shocks in the accretion torus, even for an almost Keplerian distribution of angular momentum \citep{Zanotti:2010}. In recent years there have also been several studies which included self-gravity of the disc \citep{Korobkin:2011,Kiuchi:2011,Mewes:2016}. They have shown how the non-axisymmetric structures that arise from the instability can lead to a significant emission of gravitational waves.

\cite{Hawley:2000} systematically investigated for the first time the evolution of three-dimensional magnetized tori, presenting a series of models with different gravitation potential (Newtonian and pseudo-Newtonian), distribution of angular momentum, magnetic field strength and topology (toroidal and poloidal) and azimuthal range (from one quadrant to the full $2\pi$ angle). Despite considering models unstable to the PPI, almost each model failed to display a significant growth of the PPI, as the relatively faster development of the MRI led to a suppression of the hydrodynamic instability. The only model that displayed a non-negligible (although still not dominant) growth of the PPI was the one threaded by a subthermal constant toroidal field (named CT2 in the paper). This setup triggers, in fact, a non-axisymmetric MRI mode whose growth is slower than the one selected by a vertical magnetic field, hence enabling an early linear development of the PPI.
In the last two decades several studies have investigated the dynamics of thick magnetized tori accreting onto black holes with three-dimensional global simulations \citep{Hawley:2001,Hawley:2001a,Arlt:2001b,Machida:2003,DeVilliers:2003,DeVilliers:2003b,Machida:2004,Kigure:2005,DeVilliers:2005,Fragile:2007,McKinney:2009,McKinney:2012,Wielgus:2015,Fragile:2017} and none of them reported any significant growth of the PPI once the presence of weak magnetic fields of different topologies and strenghts were taken into account.  

These results suggest that the MRI can efficiently inhibit the onset of the PPI in accretion tori. However, it is yet not completely clear what is the extent of their validity and whether there are serious limitations to them in astrophysically relevant scenarios. Indeed, the vast majority of the previously cited numerical studies considered non-equilibrium backgrounds as initial condition, since a hydrodynamic torus was threaded by an ad-hoc superimposed magnetic field. This choice inevitably leads to an initial transient in the simulation, which could in principle favour one instability over the other, and hence put in serious question the validity of the results on the relative importance of PPI and MRI. Moreover, only a very small fraction of previous works \citep{Hawley:2000,Hawley:2001,Machida:2003,Machida:2004,McKinney:2012} considered models with purely toroidal magnetic fields and the full azimuthal range, which are necessary conditions to study the interaction between the PPI and the (slow) non-axisymmetric MRI in wide tori. 
There are only a few recent works that employed analytical equilibrium solutions as initial conditions. \cite{Wielgus:2015} studied the stability of strongly magnetized tori using the analytical solution provided by \cite{Komissarov:2006}, where the extra-support of a purely toroidal magnetic field is consistently taken into account in the initial state of the disc. This setup allowed for a detailed study of the onset of non-axisymmetric MRI modes in relativistic accretion tori, but the models considered only covered a limited azimuthal range ($\phi\in\{0,\pi/2\}$). Therefore, they could only consider the dynamics of modes with an azimuthal number $m$ being a multiple of 4, filtering out a possibly dominant $m=1$ mode that would be selected for a hydrodynamical wide torus. The same limited azimuthal range was employed in \cite{Fragile:2017}, where stronger magnetic fields were also considered.
Moreover, \cite{Fu:2011} tried to analytically establish the effect of magnetic fields on the development of the PPI. Their analysis (which assumes an incompressible fluid) suggests that sufficiently strong magnetic fields can actually further destabilize the torus and reinvigorate the hydrodynamic instability. 

In this work, we present 3D GRMHD simulations of magnetized tori with the goal to establish the effect of the MRI on the PPI development starting from the equilibrium solution by \cite{Komissarov:2006}, which consistently includes the support of a purely toroidal magnetic field in the disc's initial condition. We first analyze the behaviour of an unmagnetized torus to characterize the standard development of the hydrodynamic instability. We then consider equilibria with different strengths of the toroidal field, perturbation seeds and grid resolutions to assess how the growth of non-axisymmetric global modes depends on these parameters.
For all our simulations we use the code \texttt{ECHO} \citep{Del-Zanna:2007,Bucciantini:2011}, which integrates the full set of GRMHD equations and has been used in the past to study non-Keplerian discs around black holes \citep{Zanotti:2010,Bugli:2014}. 

The plan of the paper is as follows. In Section 2 we describe the set of GRMHD equations considered in our study. The disc models and the numerical setup are provided in Section 3, while in Section 4 we introduce the diagnostics used to analyze the data produced by the simulations. We present then our results and discuss them in Section 5, and we finally give our conclusions in Section 6.   

\section{System equations}
In the following we set $c=G=M_{\odot}=1$ and absorb all factors $\sqrt{4\pi}$ in the definition of the electromagnetic field. Greek letters $\mu$, $\nu$, $\lambda$, \dots (ranging from 0 to 3) are used for 4D spacetime tensor components, while Latin letters $i$, $j$, $k$, \dots (ranging from 1 to 3) are used for 3D spatial tensor components.

The set of GRMHD equations is composed of the mass and energy-momentum conservation laws
\begin{gather}
 \nabla_\mu(\rho u^\mu)=0,  \label{eq:cons1}\\
 \nabla_\mu T^{\mu\nu}=0,\label{eq:cons2}  
\end{gather}
and Maxwell's equations
\begin{gather}
 \nabla_\mu F^{\mu\nu}=-I^{\nu},\label{eq:cov_max1}\\
 \nabla_\mu F^{*\mu\nu}=0.\label{eq:cov_max2}
\end{gather}
In the previous expressions we introduced the fluid comoving mass density $\rho$, the fluid four-velocity $u^\mu$, the stress-energy tensor $T^{\mu\nu}$, the Faraday tensor $F^{\mu\nu}$ and its dual $F^{*\mu\nu}$, and the four-vector of current density $I^\mu$. The stress-energy tensor is made up of two contributions, $T^{\mu\nu}=T_m^{\mu\nu}+T_f^{\mu\nu}$, due to the matter distribution
\be
 T_m^{\mu\nu} = \rho h u^\mu u^\nu + pg^{\mu\nu},
\ee
and the electromagnetic field
\be
 T_f^{\mu\nu} = F^\mu_{\ \ \lambda}F^{\nu\lambda}-\tfrac{1}{4}(F^{\lambda\kappa}F_{\lambda\kappa})g^{\mu\nu},
\ee
where $h=1+\tfrac{\Upsilon}{\Upsilon-1}\frac{p}{\rho}$ is the specific enthalpy, $p$ is the total pressure and $g^{\mu\nu}$ is the spacetime metric tensor. In the former definition of the enthalpy we assumed an ideal EoS for a perfect gas with adiabatic index $\Upsilon$
\be
p(\rho,\epsilon)=(\Upsilon-1)\rho\epsilon.
\ee
GRMHD equations can be decomposed according to the 3+1 formalism \citep{Baumgarte:2003}, by introducing the line element in the so-called ADM form \citep{Arnowitt:1962}
\be
ds^2=-\alpha^2\tx{d}t^2+\gamma_{ij}(\tx{d}x^i+\beta^j\tx{d}t)(\tx{d}x^j+\beta^i\tx{d}t),
\ee
where $\alpha$ is the lapse function, $\bm{\beta}$ is the shift vector, and $\gamma^{ij}$ is the spatial metric tensor. GRMHD equations can be decomposed accordingly into their spatial and temporal components by considering the frame of the \emph{Eulerian observer} (also referred to as the \emph{Zero Angular Momentum Observer}, ZAMO), i.e. the observer who moves with four-velocity
\be
n^\mu=\left(\frac{1}{\alpha},-\frac{\beta^i}{\alpha}\right).
\ee
In this frame we can decompose the 4D space-time quantities as
\begin{align}
 u^\mu &=\Gamma n^\mu+\Gamma v^\mu,\label{eq:u_decomp}\\
 T^{\mu\nu} &=W^{\mu\nu}+S^\mu n^\nu+n^\mu S^\nu+U n^\mu n^\nu,\label{eq:stress_decomp}\\
 F^{\mu\nu} &=n^\mu E^\nu-E^\mu n^\nu+\epsilon^{\mu\nu\lambda\kappa}B_\lambda n_\kappa,\label{eq:F_decomp}\\
 F^{*\mu\nu}&=n^\mu B^\nu-B^\mu n^\nu-\epsilon^{\mu\nu\lambda\kappa}E_\lambda n_\kappa,\label{eq:F*_decomp}\\
 I^\mu &=qn^\mu+J^\mu, \label{eq:I_decomp}
\end{align}
where $\epsilon^{\mu\nu\lambda\kappa}$ is the spacetime Levi-Civita tensor density and the new vector and tensors are now spatial quantities measured in the frame of the Eulerian observer. Eq. \eqref{eq:u_decomp} introduces the spatial fluid velocity $v^\mu$ of Lorentz factor $\Gamma$, given by
\begin{align}
 v^i &=\frac{u_i}{\Gamma}+\frac{\beta^i}{\alpha},\\
 \Gamma &=\alpha u^t=(1-v_i v^i)^{-1/2}.
\end{align}
In Eq. \eqref{eq:stress_decomp} appear the spatial stress-tensor $W^{\mu\nu}$, the momentum density $S^\mu$, and the energy density $U$, while in Eq. \eqref{eq:F_decomp} and \eqref{eq:F*_decomp} we find the spatial electromagnetic vectors $E^\mu=-n_\nu F^{\nu\mu}$ and $B^\mu=-n_\nu F^{*\nu\mu}$. Finally, $q$ and $J^\mu$ in Eq. \ref{eq:I_decomp} represent the electric charge density and the spatial conduction current, respectively. Combining Eqs. \eqref{eq:u_decomp}-\eqref{eq:F*_decomp} we can also express $W^{\mu\nu}$, $S^\mu$ and $U$ in terms of the other spatial quantities as
\begin{align}
 W^{ij}&=\rho h\Gamma^2 v^iv^j-E^iE^j-B^iB^j+\left[p+\tfrac{1}{2}\left(E^2+B^2\right)\right]\gamma^{ij}, \\
 S^i&=\rho h\Gamma^2 v^i+\epsilon^{ijk}E_jB_k,\label{eq:mom_dens} \\
 U&=\rho h\Gamma^2-p+\tfrac{1}{2}\left(E^2+B^2\right),\label{eq:en_dens}
\end{align}
where $\epsilon^{ijk}=-n_\mu\epsilon^{\mu ijk}$ is the spatial Levi-Civita tensor density.
Eqs. \eqref{eq:cons1} and \eqref{eq:cons2} can be cast into the following compact conservative form 
\begin{equation}\label{eq:conserv_law}
 \partial_{t} \bm{\mathcal{U}}+\partial_{j} \bm{\mathcal{F}}^j=\bm{\mathcal{S}},
\end{equation}
where we introduced the conservative variables $\bm{\mathcal{U}}$, the fluxes $\bm{\mathcal{F}}^j$, and the source terms $\bm{\mathcal{S}}$, defined respectively as  
\begin{gather}\label{u}
 \bm{\mathcal{U}}=\gamma^{1/2}\left[\begin{array}{c}
                                    D \\S_i\\ U 
                                    \end{array}\right], \\
 \label{f}
 \bm{\mathcal{F}}^j=\gamma^{1/2}\left[\begin{array}{c}
                                    \alpha v^jD-\beta^jD \\\alpha W^j_i-\beta^jS_i\\ \alpha S^j-\beta^jU\\ 
                                    \end{array}\right],\\
 \label{s}
 \bm{\mathcal{S}}=\gamma^{1/2}\left[\begin{array}{c}
                                    0 \\
                                    \tfrac{1}{2}\alpha W^{jk}\partial_i\gamma_{jk}+S_j\partial_i\beta^j-U\partial_i\alpha\\ \tfrac{1}{2} W^{jk}\beta^i\partial_i\gamma_{jk}+W_j^i\partial_i\beta^j-S^i\partial_i\alpha\\
                                    \end{array}\right],
\end{gather}
where $D=\Gamma\rho$ is the mass density measured by the Eulerian observer and $\gamma$ is the determinant of the spatial metric tensor. The expression for the source terms in Eq. \eqref{s} is valid only in the case of a stationary metric, i.e. in the Cowling approximation.

We then add to Eqs. \eqref{eq:conserv_law} the 3+1 decomposition of Maxwell's equations \eqref{eq:cov_max1} and \eqref{eq:cov_max2} for the evolution of the electric and magnetic fields 
\begin{gather}
 \partial_{t} \left(\gamma^{1/2}B^i\right)+\gamma^{1/2}\epsilon^{ijk}\partial_{j}\left(\alpha E_k+\epsilon_{klm}\beta^l B^m\right)=0, \\
 \begin{split}
 \partial_{t} \left(\gamma^{1/2}E^i\right)+\gamma^{1/2}\epsilon^{ijk}\partial_{j}\left(-\alpha B_k+\epsilon_{klm}\beta^l E^m\right)=\\ -(\alpha J^i-q\beta^i),
 \end{split}
\end{gather}
Although the \texttt{ECHO} code can use a covariant closure for Ohm's law that includes magnetic resistivity and mean-field dynamo action (see \cite{Bucciantini:2013,Del-Zanna:2014} for more details on the modelling, \cite{Bugli:2014} for a study of $\alpha-\Omega$ dynamos in accretion discs, and \cite{Del-Zanna:2016} for an application to relativistic reconnection), here we assume the plasma to be a perfect conductor by using the condition of vanishing electric field in the fluid comoving frame:
\begin{equation}
 F^{\mu\nu}u_\nu=0 \Rightarrow \bm{E}=-\bm{v}\times\bm{B}. 
\end{equation}
Therefore, we regard the electric field as a derived quantity and neglect Ampere's law as an evolution equation. The effects of magnetic dissipation will be included in a forthcoming work.

In this work we investigate the time evolution of the following \emph{primitive} variables:
\begin{equation}\label{prim}
 \bm{\mathcal{V}}=[\rho,v^j,p,B^j],
\end{equation}
while our code \texttt{ECHO} evolves in time the conservative variables defined in Eq. \eqref{u} and the magnetic field $\bm{B}$. The inversion from conservative to primitive variables is performed by using a multidimensional Newton-Raphson scheme, which retrieves the correct value of the fluid velocity $\bm{v}$ from Eq. \eqref{eq:mom_dens} combined with the definitions of $D=\Gamma\rho$ and $U$ in Eq. \eqref{eq:en_dens}.

\begin{table}\centering
\caption{List of the parameters defining the initial unperturbed state of the hydrodynamic torus. The same parameters were used to initialize the magnetized models (see \reftab{tab:hd-models} for the additional ones related to the magnetic field).}\label{tab:disc-models}
\centering
\begin{tabular}{cccccc}
\toprule
\toprule
$M_\tx{BH}$ & $a$ & $r_\tx{in}$ & $r_\tx{c}$ & $l$ & $\rho_c$ \\ 
\midrule
1 & 0 & 6.16 & 10.17 & 3.97 & 1 \\
\bottomrule
\bottomrule
\end{tabular}
\end{table}

\section{Disc model and numerical setup}\label{sec:simulation}
Since we are interested in the dynamics taking place within the torus and not in proximity of the black hole event horizon, we consider for the sake of simplicity a non-rotating black hole of mass $M=1$ in a spherical coordinate system $(r,\theta,\phi)$. The black hole is surrounded by a thick torus \citep{Abramowicz:1978} whose inner edge and center are located at $r_\tx{in}=6.16$ (close to the last marginally stable orbit located at $r_\tx{ms}=6$) and $r_\tx{c}=10.17$, respectively. This choice sets the specific angular momentum of the disc to $l=3.97$. The orbital period at the disc center is $P_c\sim207$. We adopt the Cowling approximation, i.e. assuming a time independent metric, and neglect the self gravity of the disc, disregarding also any change in the central black hole mass and spin due to accretion. Density can therefore be rescaled to the central peak value at $r_c$, that is $\rho_c$ (in the following density is always to be intended in this way). The adiabatic index is set everywhere to $\Upsilon=4/3$.

For the magnetized tori we initialize our simulations with the stationary solution provided by \cite{Komissarov:2006}, keeping the same parameters as in the hydrodynamic models for black hole spin, disc's inner edge and center location and density normalization (see \reftab{tab:disc-models}). We vary the value of the magnetization $\sigma=B^2/2p$ at the disk's center, i.e. $\sigma_c$, to investigate the role of the magnetic field strength on the system's stability. Further details on the initialization of the magnetized thick disc can be found in \cite{Del-Zanna:2007}.

The atmosphere is initialized as a Michel's radial inflow \citep{Michel:1972}, a stationary solution in the Schwarzschild metric determined by the  adiabatic index $\Upsilon=4/3$ and the value of the atmospheric density at distance $r_\tx{c}$, i.e. $\rho_\tx{atm}=10^{-6}$. To provide stability for the integration, we set a numerical floor value for the density equal to $\rho_\tx{fl}=10^{-9}$. 

We adopt Kerr-Schild coordinates (see Appendix A) to allow for an inner radial boundary inside the black hole event horizon located at a radius $r_h=2$, thus preventing numerical artifacts due to boundary effects that could otherwise propagate through the domain and affect the simulation at $r>r_h$. In radial direction the numerical domain ranges from $r_\tx{min}=0.97~r_h$ to $r_\tx{max}=100$ covered by $N_r=256$ grid points, with outflow boundary conditions ($0^\tx{th}$ order extrapolation) applied at both radial extrema. The radial mesh is non-uniform to increase the resolution towards the black hole event horizon by defining our radial grid points $r_i$ as
\be
  r_i=r_\tx{min}+\frac{r_\tx{max}-r_\tx{min}}{\varepsilon}\tan\left[\arctan(\varepsilon)x_i\right],
\ee
where $x_i=(i-0.5)/N_r$ and the stretching parameter $\varepsilon$ is set to 10. 
The polar domain extends from $\theta_\tx{min}=0$ to $\theta_\tx{max}=\pi$ with a resolution of $N_\theta=256$ points. For simplicity we impose axisymmetric reflection at the boundaries, since we do not expect our three-dimensional models to be affected by the dynamics in the low density regions close to the rotation axis. To better resolve the disc, the polar mesh is refined towards the equatorial midplane by setting the polar grid points $\theta_i$ to
\be
  \theta_i=\frac{\pi}{2}[1+(1-\zeta)(2y_i-1)+\zeta(2y_i-1)^n],
\ee
where $y_i=(i-0.5)/N_\theta$, $\zeta=0.6$ and $n=29$. This gives a roughly constant and fine grid spacing across the disc and a rapidly decreasing resolution towards the rotational axis \citep{Noble:2010}. Finally, we consider the full azimuthal range $\phi\in[0, 2\pi]$ with uniformly distributed cells and periodic boundaries to be able to resolve global azimuthal modes with mode number $m=k_\phi r=1$, which are expected to develop and to be also the fastest growing modes for the PPI in our disc model. 

An important aspect to consider in any numerical experiment is its convergence, i.e. whether or not the results depend on the grid resolution. 
For a simulation involving magnetized accretion flows the key aspect that needs to be properly resolved is the MRI turbulence that appears whenever a differentially rotating fluid is threaded by a magnetic field of any topology. Following \citet{Hawley:2011} we define a quality metric as the ratio of the characteristic wavelength of the MRI mode $\lambda_{\tx{MRI}}=2\pi |u_{\tx{A}}|/\Omega$ (which corresponds to the distance travelled by an Alfv\'en wave during an orbital period) and the grid zone size. Since we start with a purely toroidal magnetic field, the relevant quality metric should consider the wavelength along the $\phi$ direction, that is
\begin{align}
 Q_{\phi}=\frac{\lambda_{\tx{MRI}}}{\Delta \phi\sqrt{\gamma_{\phi\phi}}}=\frac{2\pi|u_{\tx{A}\phi}|}{\Omega\Delta \phi\sqrt{\gamma_{\phi\phi}}}.
\end{align}
The volume-average of $Q_\phi$ for each magnetized model at the beginning of the simulation is reported in \reftab{tab:hd-models}. \cite{Hawley:2011} suggest that $Q_{\phi}\gtrsim 20$ should provide a sufficiently good description of the non-linear phase of MHD turbulence. Note, however, that the recent stratified shearing box simulations of \cite{Ryan:2017} suggest that none of the current simulations may actually be converged, even at much higher resolution than achievable in a global model. For all our simulations we use the Harten-Lax-van Leer Riemann-solver instead of the more dissipative Lax-Friedrichs scheme, together with a PPM reconstrution scheme. 

To trigger the growth of non-axisymmetric modes we introduce inside the torus a small perturbation $\delta v^\phi$ of the equilibrium azimuthal velocity $v^\phi_0$, with either random noise or cosine waves of the form:
\be
  \delta v^\phi = A v^\phi_0\cos(m\phi),
\ee
with $m=1,...,5$ and amplitudes $A$ ranging from $10^{-6}$ to $10^{-2}$ depending on the simulation.

\begin{table*}\centering
\caption{List of the models considered in our study. $N_r$, $N_\theta$ and $N_\phi$ are the number of grid points in radial, polar and azimuthal direction, and $A(v^\phi)$ is the amplitude of the initial perturbations. $\sigma_c$ and $|\bm{B}|_c$ are respectively the initial value of the magnetization and magnetic field magnitude at the center of the disc. $\langle Q_\phi\rangle_V$ is the initial volume-average of the quality metric $Q_\phi$ (weighted by the rest mass density $\rho$).}\label{tab:hd-models}
\centering
\begin{tabular}{cccccccccc}
\toprule
\toprule
& $N_r$ & $N_\theta$ & $N_\phi$ & $A(v^\phi)$ & Excitation & $\sigma_c$ & $|\bm{B}|_c$ & $\langle Q_\phi\rangle_V$\\ 
\midrule
H64r3 & 256 & 256 & 64 & $10^{-3}$ & Random  & 0 & 0 & 0 \\
H256r-4 & 256 & 256 & 256 & $10^{-4}$ & Random & 0 & 0 & 0 \\
H32m1-4 & 256 & 256 & 32 & $10^{-4}$ & $m=1$ & 0 & 0 & 0 \\
H64m1-4 & 256 & 256 & 64 & $10^{-4}$ & $m=1$ & 0 & 0 & 0 \\
H128m1-4 & 256 & 256 & 128 & $10^{-4}$ & $m=1$ & 0 & 0 & 0 \\
H256m1-4 & 256 & 256 & 256 & $10^{-4}$ & $m=1$ & 0 & 0 & 0 \\
H64m2-2 & 256 & 256 & 64 & $10^{-2}$ & $m=2$ & 0 & 0 & 0 \\
H64m3-2 & 256 & 256 & 64 & $10^{-2}$ & $m=3$ & 0 & 0 & 0 \\
H64m4-2 & 256 & 256 & 64 & $10^{-2}$ & $m=4$ & 0 & 0 & 0 \\
H64m5-2 & 256 & 256 & 64 & $10^{-2}$ & $m=5$ & 0 & 0 & 0 \\
\midrule
\modela    & 256 & 256 & 256 & $10^{-3}$ & Random & 0                 & 0                   & 0 \\
\modelb    & 256 & 256 & 256 & $10^{-3}$ & Random & $10^{-2}$         & $8.06\times10^{-3}$ & 6.94\\
\modelB    & 256 & 256 & 512 & $10^{-3}$ & Random & $10^{-2}$         & $8.06\times10^{-3}$ & 13.89\\
\modelc    & 256 & 256 & 256 & $10^{-3}$ & Random & $3\times 10^{-2}$ & $1.38\times10^{-2}$ & 11.88\\
\modelC    & 256 & 256 & 512 & $10^{-3}$ & Random & $3\times 10^{-2}$ & $1.38\times10^{-2}$ & 23.75\\
\modeld    & 256 & 256 & 256 & $10^{-3}$ & Random & $10^{-1}$         & $2.43\times10^{-2}$ & 20.79\\
\modelD    & 256 & 256 & 512 & $10^{-3}$ & Random & $10^{-1}$         & $2.43\times10^{-2}$ & 41.57\\
\modelA    & 256 & 256 & 512 & $10^{-3}$ & $m=1$  & 0                 & 0                   & 0 \\
\modelBm   & 256 & 256 & 512 & $10^{-3}$ & $m=1$  & $10^{-2}$         & $8.06\times10^{-3}$ & 13.89\\
\bottomrule
\bottomrule
\end{tabular}
\end{table*}

\section{Diagnostics}
We now introduce the quantities that we calculate from each simulation to probe the dynamical evolution of the models and determine the relative importance of the PPI and MRI in these models.

\subsection{Power of azimuthal modes}
In analogy with \cite{De-Villiers:2002} and \cite{Wielgus:2015}, for any given azimuthal number $m$ we calculate the power contained in an azimuthal mode for a generic quantity $Q$ as
\be\label{eq:mode_power}
\mathcal{P}_{m,Q}(t) = \frac{\int_{r_\tx{in}}^{r_\tx{out}}\int_{0}^{\pi}\left| \frac{1}{2\pi}\int_{0}^{2\pi}Q e^{im\phi}\tx{d}\phi\right|^2 w(r,\theta)\sqrt{\gamma}\tx{d}\theta\tx{d}r}{\int_{r_\tx{in}}^{r_\tx{out}}\int_{0}^{\pi}w(r,\theta)\sqrt{\gamma}\tx{d}\theta\tx{d}r},
\ee
where $w(r,\theta)$ is an axisymmetric weight-function. For $Q\in\{u^\phi,u^\phi_A\}$ (where $u_{\tx{A}}=B/\sqrt{\rho h+B^2}$ is the relativistic Alfv\'en velocity) the integrals in radial and polar direction in Eq.~(\ref{eq:mode_power}) are weighted by the rest mass density, i.e. 
\be
w=\langle\rho\rangle_\phi=\frac{1}{2\pi}\int_0^{2\pi}\rho~\tx{d}\phi,
\ee
where the operator $\langle\rangle_\phi$ represents an azimuthal average. This choice avoids overestimating the contribution of the rarefied atmosphere enveloping the disc and at the same time computing quantities that relate respectively to the azimuthal components of kinetic energy and magnetic energy. For $Q=\rho$ we set $w=1$.
Since we are interested mostly in the relative importance of the $m=1$ mode with respect to other higher order non-axisymmetric modes, we decided to normalize the mode power in Eq.~(\ref{eq:mode_power}) at any given time by $\mathcal{P}_{0,Q}(t)$, i.e. by the instantaneous power in the axisymmetric mode. By doing so, we factor out any decrease of the azimuthal power introduced by the significant mass loss occuring in all the models we considered. The evident drawback of this time-dependent normalization is a loss of information in absolute terms on the evolution of the power within a specific azimuthal mode, once a non-negligible fraction of the disc's mass has been accreted onto the black hole. However, we verified that this normalization did not affect the estimate of growth rates and the value at which the modes power saturates, since most mass is accreted during later stages.

With these diagnostics we estimate growth rates and saturation levels. We also quantify the time evolution of the relative importance of the various modes resolved by the numerical simulations by constructing spectrograms. At any given time $t$, we compute the power in Eq.~(\ref{eq:mode_power}) of modes with azimuthal number up to $m=50$, and we plot the power in a $m$ vs. time diagram. By time-averaging over the full duration of the simulations, we also compute spectra to characterize the power distribution across different modes.

We display information on the frequency components present in the fastest growing PPI mode with frequency-radius diagrams. We consider the complex amplitude of the $m=1$ mode of the density in the equatorial plane (since most of the dynamics takes place in this region):
\be
\mathcal{M}(r,t)=\frac{1}{2\pi}\int_{0}^{2\pi}\rho(r,\pi/2,\phi)e^{i\phi}\tx{d}\phi,
\ee
For each radius $r$ we compute $\tilde{\mathcal{M}}(r,\omega)=FFT(\mathcal{M}(r,t))$, i.e. the Fourier Transform in the frequency domain.

\subsection{Turbulence and accretion}
We keep track of the development of turbulence in the system by considering the evolution of the turbulent kinetic energy density, defined as the difference between the total kinetic energy density and the component due to the mean orbital motion of the fluid. Integrating this quantity over the computational volume $V$ we obtain:
\be
K_{\tx{turb}}=\frac{1}{2}\int\limits_V \rho\delta u^2 \,\tx{d}V,
\ee
where $\delta u^2=u^ru_r+u^\theta u_\theta+(u^{\phi}-\langle u^\phi\rangle_\phi)^2\gamma_{\phi\phi}$ and $\tx{d}V=\sqrt{\gamma}\,\tx{d}r\,\tx{d}\theta\,\tx{d}\phi$ is the covariant volume element.

Another set of quantities useful to probe the dynamical evolution of the system are the $r-\phi$ components of the Reynolds and Maxwell stress tensors, defined respectively as:
\begin{align}
W_{\tx{Re}}&= \rho\,\delta u^r\,\delta u^\phi\sqrt{\gamma_{rr}}\sqrt{\gamma_{\phi\phi}}, \\
W_{\tx{Ma}}&= B^r\,B^\phi\,\sqrt{\gamma_{rr}}\,\sqrt{\gamma_{\phi\phi}},
\end{align}
where $\delta u^r=u^r-\langle u^r\rangle_\phi$ and $\delta u^r=u^r-\langle u^r\rangle_\phi$. We compute their volume averages by considering only those regions of the computational domain where the rest mass density $\rho$ exceeds a threshold value set to $\rho_{\tx{th}}=\sqrt{\rho_{\tx{c}}\,\rho_{\tx{atm}}}$ to track the dynamics of the disc and exclude that of the atmosphere.

Still related to the stresses, we compute the disc \emph{alpha} parameter (not to be confused with the lapse function) as the ratio of the volume average of the total stress $W_{\tx{tot}}=W_{\tx{Re}}+W_{\tx{Ma}}$ and the volume average of the thermal pressure:
\be
\alpha_\tx{turb}=\frac{\langle W_{\tx{tot}}\rangle_V}{\langle p\rangle_V}.
\ee
As a further diagnostics of the efficiency of angular momentum transport in the disc, and thus of the overall accretion process, we also monitor the evolution of the radial distribution of the disc's orbital angular velocity $\Omega$. The radial dependence of $\Omega$ is usually described with a power-law:
\be\label{eq:slope_omega}
\Omega\propto r^{-q},
\ee
where the parameter $q$ can range from 3/2 (for a Keplerian disc) to 2 (constant specific angular momentum). However, in the relativistic case and for a non-rotating black hole $\Omega=-lg_{tt}/g_{\phi\phi}=l(r-2)/r^3$, and the  value $q=2$ can be assumed only if the disc extends sufficiently far away from the black hole. Since our disc model extends from $r_\tx{in}=6.16$ to $r_\tx{out}=21.6$, initially q ranges from $q(r_\tx{in})=1.52$ to $q(r_\tx{out})=1.90$. Since we want to relate the redistribution of specific angular momentum to the classical case where the slope in Eq.~(\ref{eq:slope_omega}) starts as a constant equal to 2 across the disk, instead of $q$ we monitor the evolution of the quantity
\be\label{eq:tilde_q}
\tilde{q}=2-\left|\frac{\tx{d}\log l}{\tx{d}\log r}\right|,
\ee
which is evaluated performing a least-squares fit of the power law describing the radial dependence of the specific angular momentum $l$.

\section{Results and discussion}\label{sec:conc}
\begin{figure}\centering
     \includegraphics[scale=0.4]{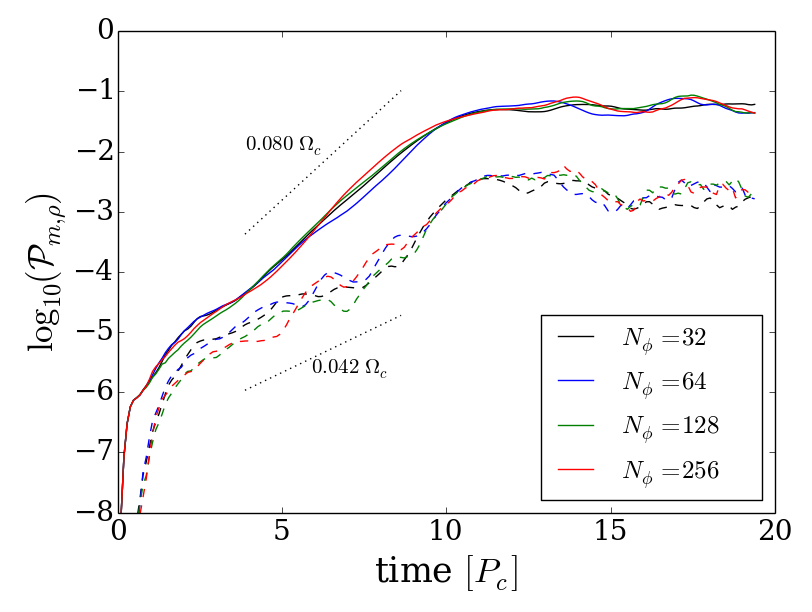}
    \caption{Time evolution of the power in density for the $m=1$ (solid curves) and $m=2$ (dashed curves) modes, as computed from Eq.~(\ref{eq:mode_power}) for models H32m1-4 to H256m1-4. All models are initialized with an $m=1$ perturbation.}
    \label{power_resolution}
\end{figure}

  \subsection{Hydrodynamic disc}
  We first focus on the development and saturation of the PPI in the absence of magnetic fields to have an initial benchmark for a later comparison with the results for magnetized models. As shown in \reftab{tab:hd-models}, we performed a set of simulations that differ by resolution in the azimuthal direction, and amplitude and spectrum of the initial perturbation.
  We followed the evolution of these models up to 20 orbital periods at the disc center, which is sufficient for the hydrodynamic instability to reach saturation in terms of azimuthal mode power.
  
  Our results confirm that the $m=1$ azimuthal mode is the fastest growing one. Therefore, it was selected by the system independently of the initial perturbation.
  We ran a series of simulations with the same monochromatic $m=1$ perturbation but with different resolutions in the azimuthal direction (models H32m1-4 to H256m1-4). \refig{power_resolution} shows the time-evolution of the azimuthal power of the density fluctuations for the $m=1$ and $m=2$ modes. Even with a modest resolution of 32 zones, \texttt{ECHO} is capable of capturing the dynamical evolution of the PPI, since the most unstable mode has a quite large wavelength. \refig{power_resolution} also shows the linear phase of the instability during the first ten orbital periods, with growth rates for the $m=1$ and $m=2$ modes of $0.080\ \Omega_c$ and $0.042\ \Omega_c$, respectively. The power of the two modes differs also in the value of saturation levels, which is more than an order of magnitude larger for the $m=1$ mode.
  
  The minimum resolution required to properly resolve the fastest growing mode of the PPI can be much higher in radial direction than in azimuthal direction (i.e. higher than 32 points), because the $m=1$ mode developing from wide tori is not (as in the case of slender tori) the so-called \emph{principal mode}. This mode is the result of the interaction of two node-less (across the radius) surface gravity waves (also referred to as \emph{edge-waves}) that are generated at the disc's inner and outer boundaries and are advected by the shear flow with respect to the corotation radius. While this mode is essentially incompressible, the one that we see in our simulations is the outcome of the interaction between a pressure wave in the outer part of the disc and an edge wave from the interior. The former has multiple nodes in radial direction (two, in the case of our models), which require an adequate radial resolution. If one has insufficient resolution, one systematically underestimates the instability growth rates (see \cite{Blaes:1988} for a detailed discussion).
  
\begin{figure*}
    \includegraphics[trim={0 0 140 37},clip,scale=0.49]{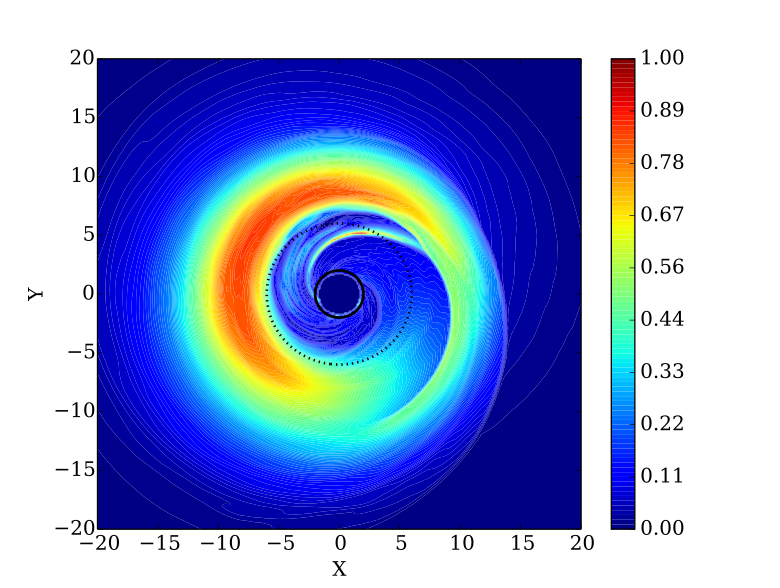}
    \includegraphics[trim={39 0 0 37},clip,scale=0.49]{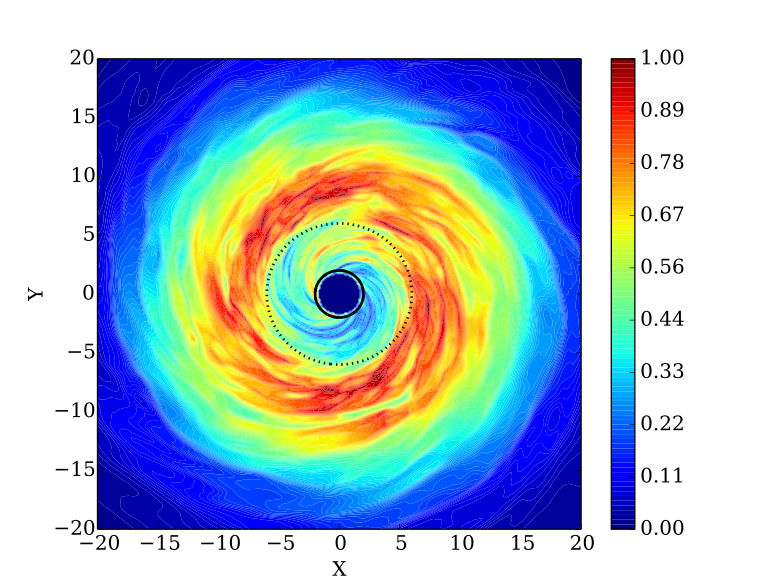}
    \includegraphics[trim={0 0 140 37},clip,scale=0.49]{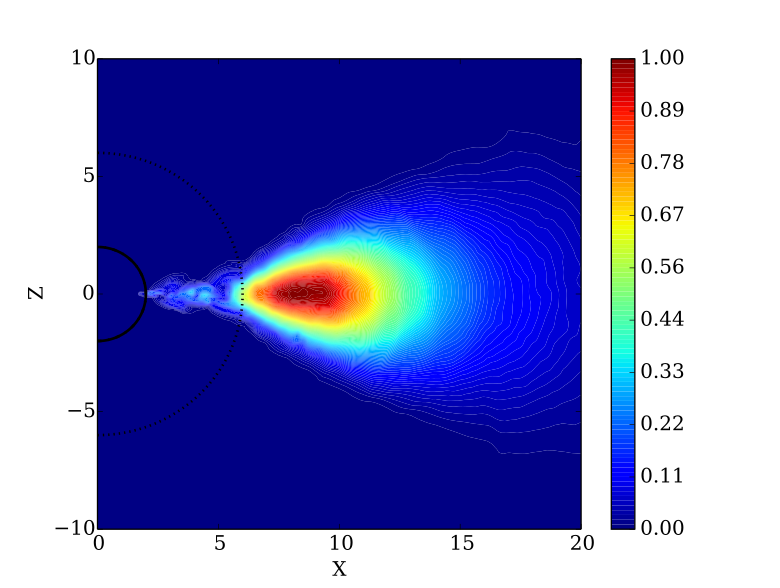}
    \includegraphics[trim={39 0 0 37},clip,scale=0.49]{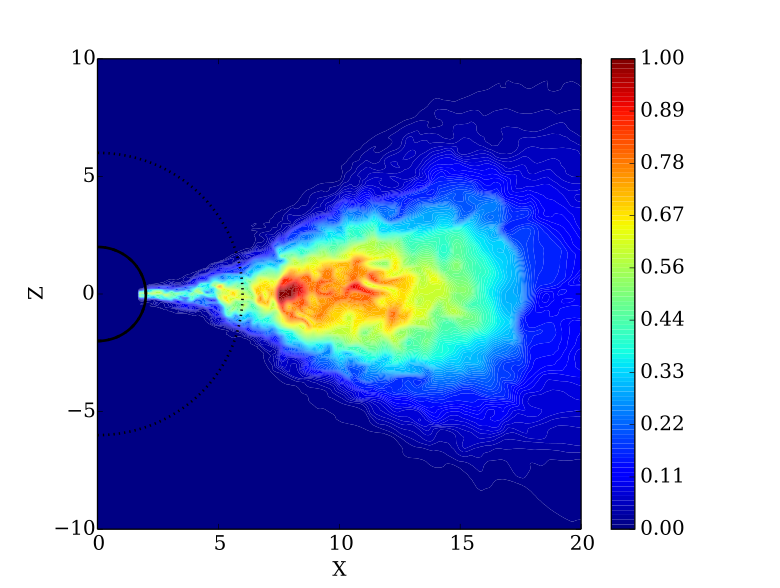}
\caption{Equatorial (top) and meridional (bottom) cuts of the rest mass density $\rho$ for models \modela~(left) and \modelB~(right) at $t=3000\simeq15\ P_c$. The maximum value of $\rho$ is normalized to 1 in each plot. The solid black curve represents the black hole event horizon, while the dotted curve indicates the radius of the last marginally stable orbit $r_\tx{ms}$.\label{fig:slices}}
\end{figure*}

  \refig{fig:slices} (top left panel) shows an equatorial slice of the rest mass density for model \modela\ after about 15 orbital periods. The dominant $m=1$ mode is clearly visible as an overdensity that corotates with the disc, while the flow still maintains overall a smooth profile. The region between the black hole horizon and the disc's inner edge is relatively depleted of mass, apart from an inspiraling flow that detached from the main body of the disc. 
  
  From the time evolution of the mode power in \refig{fig:power_rho} and the spectrogram in \refig{fig:spectrogram_rho} (top panel) it is clear that the $m=1$ mode dominates since very early times, and no other small-scale perturbation grows as much during the linear phase of the instability. After 10 orbital periods, a further deposition of energy occurs on smaller scales presumably because of non-linear interactions, but the $m=1$ mode remains the strongest one. This interpretation is confirmed by the density power spectrum (top panel of \refig{fig:spectra}), which shows also most power in low order modes, peaking at $m=1$, and decaying as a power-law $m^{-4}$ for $m\lesssim 3$. 

  \subsection{Magnetized discs}
 We now consider the effect of a weak toroidal magnetic field. In order to avoid transients, we chose the analytical solution provided by \cite{Komissarov:2006} instead of superimposing a magnetic field to the hydrodynamic solution. The drawback of such an approach is that the profile of gas density and pressure depends on the strength of the initial magnetic field. The initial magnetization is chosen small enough such that differences in these quantities between initially magnetized models and unmagnetized ones never exceeded a few percents in gas density and pressure for all our models. We therefore consider the differences to be dynamically insignificant. Moreover, recent models by \cite{Fragile:2017} suggest that strongly magnetized thick discs with a purely toroidal magnetic field may experience a fast drop of magnetization, mostly due to redistribution of gas and migration of magnetic fields into the funnel region. As in this work we are more interested in the dynamics taking place within the torus, we hence limit ourselves to values of central magnetization up to $\sigma_c=0.1$. 
 
 We focus for the moment on the magnetized models with the highest resolution along the $\phi$ direction (i.e. 512 grid points), leaving the discussion of the effects of a lower resolution to the next paragraph. As illustrated by the right panels of \refig{fig:slices}, the distribution of rest mass density in the magnetized models show small-scale fluctuations not present in the hydrodynamic models, with the absence of a self-evident large-scale overdensity. The flow consists of much smaller scales in the magnetic models, and both the equatorial and meridional slices indicate MHD turbulence triggered by the MRI.
 
\begin{figure}\centering
    \includegraphics[scale=0.42]{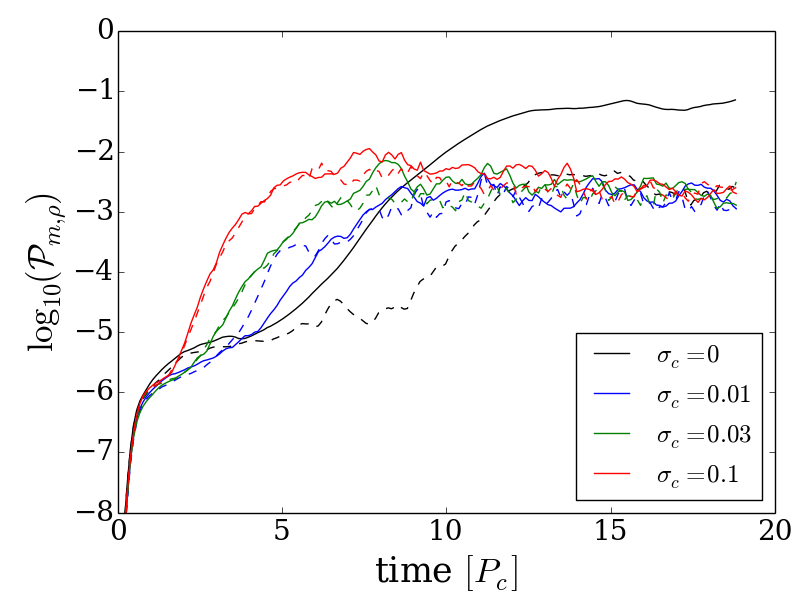}
    \caption{Time evolution of the power in density for the $m=1$ (solid curves) and $m=2$ (dashed curves) modes for models \modela~(black), \modelB~(blue), \modelC~(green), and \modelD~(red). All models are initialized with a random perturbation.}
    \label{fig:power_rho}
\end{figure}

\begin{figure}\centering
  \includegraphics[trim={1090 0 0 90},clip,scale=0.42]{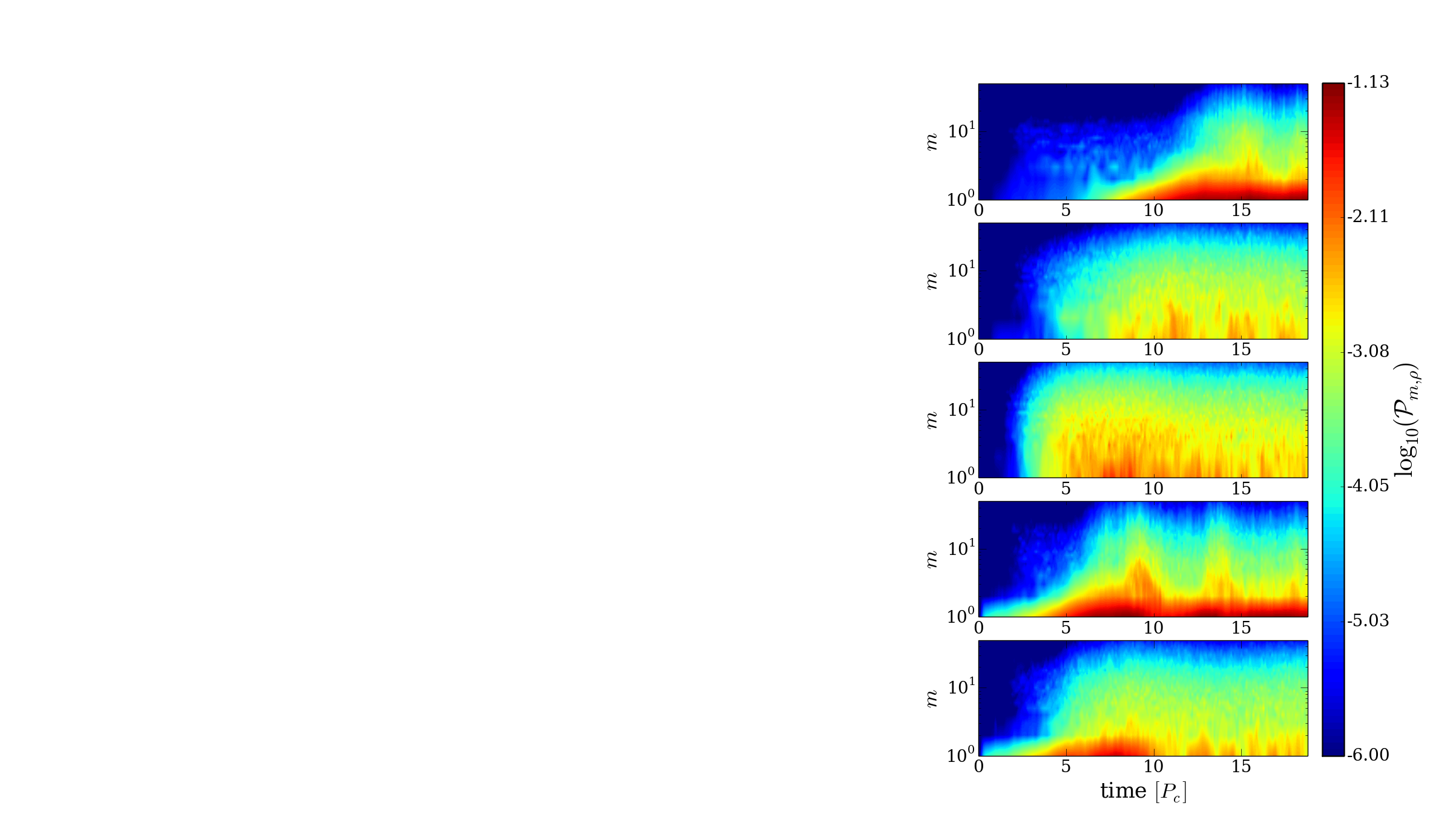}
  \caption{Rest mass density spectrograms of the azimuthal mode power, colour-coded in logarithmic scale. The panels refer, from top to bottom, respectively to models \modela,  \modelB, \modelD, \modelA~and \modelBm.\label{fig:spectrogram_rho}}  
 \end{figure}

 The time-evolution of the azimuthal mode power in the magnetized models (\refig{fig:power_rho}, blue, green and red curves with increasing magnetic field strength) reveals an earlier growth of low order modes, without a clear distinction between the $m=1$ and $m=2$ modes as in the hydrodynamic case.
 The mode power reaches a maximum, which occurs at earlier times and at a slightly higher value the stronger the magnetic field strength, although both the $m=1$ and $m=2$ modes still saturate at roughly the same level. This behaviour is confirmed also by the spectrograms in \refig{fig:spectrogram_rho}, that show a much broader range of excited modes in the magnetized cases (second and third from the top panels).
 
\begin{figure}\centering
    \includegraphics[scale=0.42]{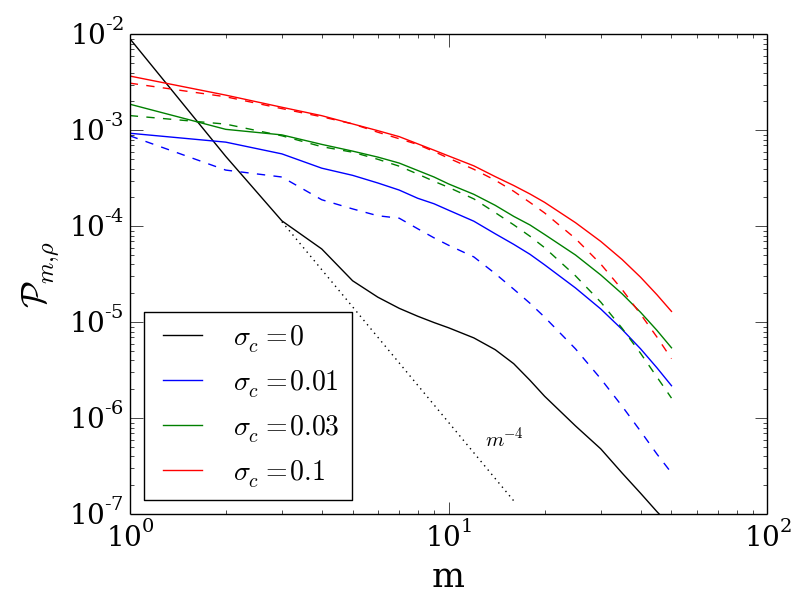}
    \includegraphics[scale=0.42]{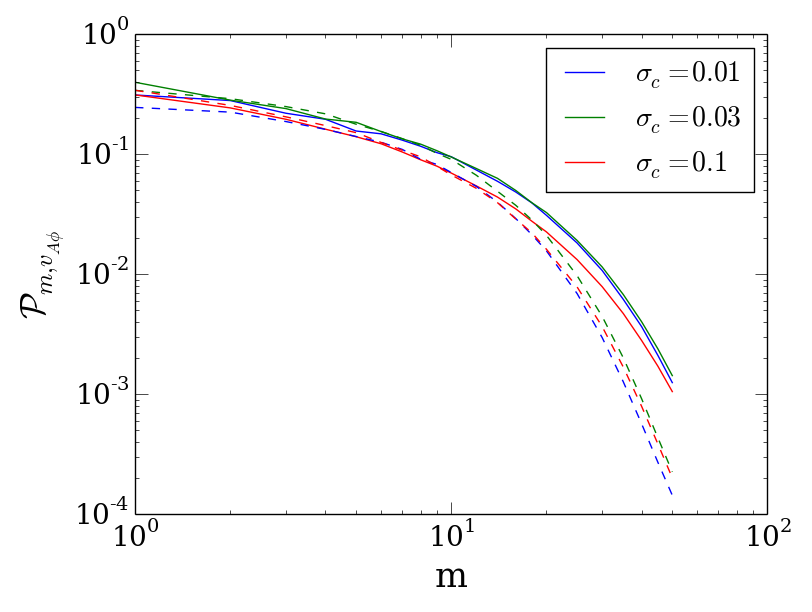}
    \caption{Time-averaged rest mass density (top) and azimuthal Alfv\'en velocity (bottom) spectra in azimuthal number $m$. The black curve represents the hydrodynamic model, the other solid curves refer to magnetized models with high resolution along the $\phi$ direction (i.e. \modelB, \modelC~and \modelD), while the dashed ones stand for the low-resolution models \modelb, \modelc, and \modeld.}
    \label{fig:spectra}
\end{figure}

 The time-averaged density spectra shown in the top panel of \refig{fig:spectra} provide a more quantitative confirmation of this trend. The hydrodynamic model \modela~(black curve) develops predominantly large-scale modes, with a steep power-law decline from the $m=1$ mode to $m\approx5$, followed by a shallower decline up to $m=10$ and again a steep drop. The magnetized models (coloured curves) have no strong excess of power at large scales, with a much shallower slope down to $m=10$ followed by a steeper decrease due to the numerical dissipation introduced by the reconstruction scheme. The spectra of the high-resolution models (coloured solid curves) behave quite similarly to each other. For $\sigma_c=0.01,0.03$ the spectra computed from the orbital Alfv\'en velocity show no particular difference in their shape (bottom panel of \refig{fig:spectra}), as they overlap quite well for both small and large azimuthal numbers. The high-magnetization model ($\sigma_c=0.1$) exhibits a slightly steeper profile and also less power, which is probably due to the stronger advection of magnetic flux towards the black hole. 
 
\begin{figure}\centering
    \includegraphics[width=0.45\textwidth]{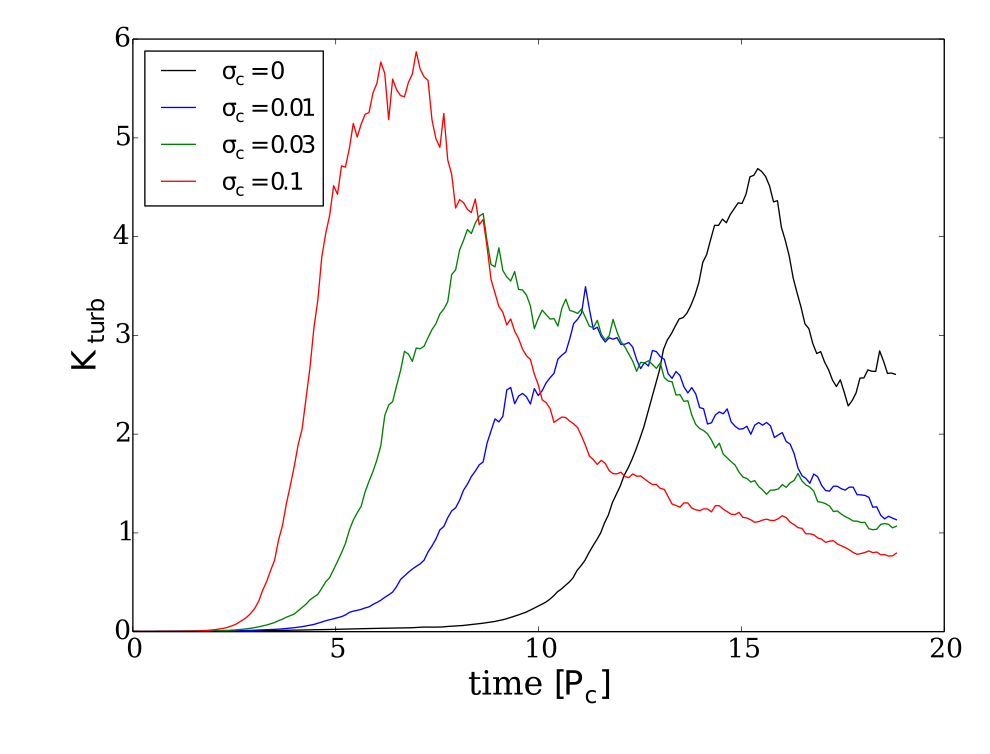}
    \includegraphics[width=0.45\textwidth]{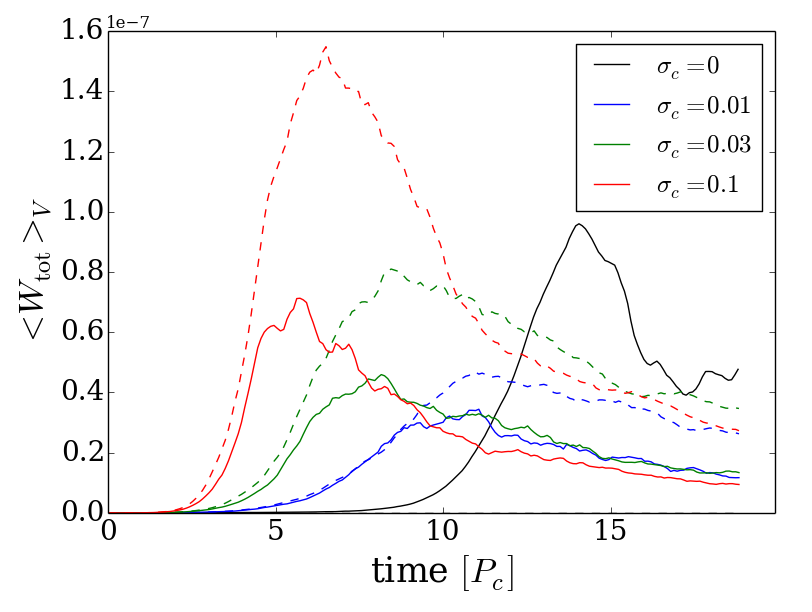}
     \includegraphics[width=0.45\textwidth]{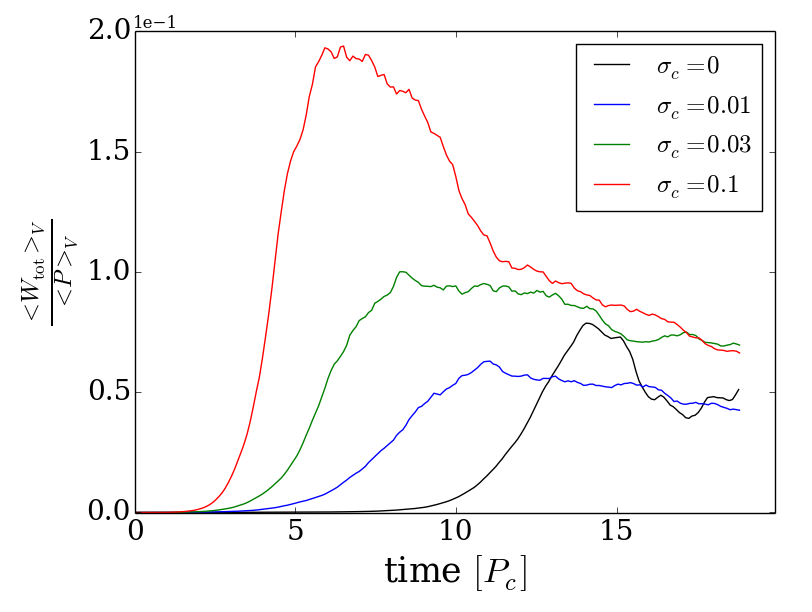}
    \caption{Turbulent kinetic energy (upper panel), stresses (midddle panel) and $\alpha_\tx{turb}$ (lower panel) for models \modela, \modelB, \modelC, and \modelD~(black, blue, green, and red as in the previous figure).}
    \label{fig:energy_stress}
\end{figure}

\begin{figure}\centering
    \includegraphics[width=0.49\textwidth]{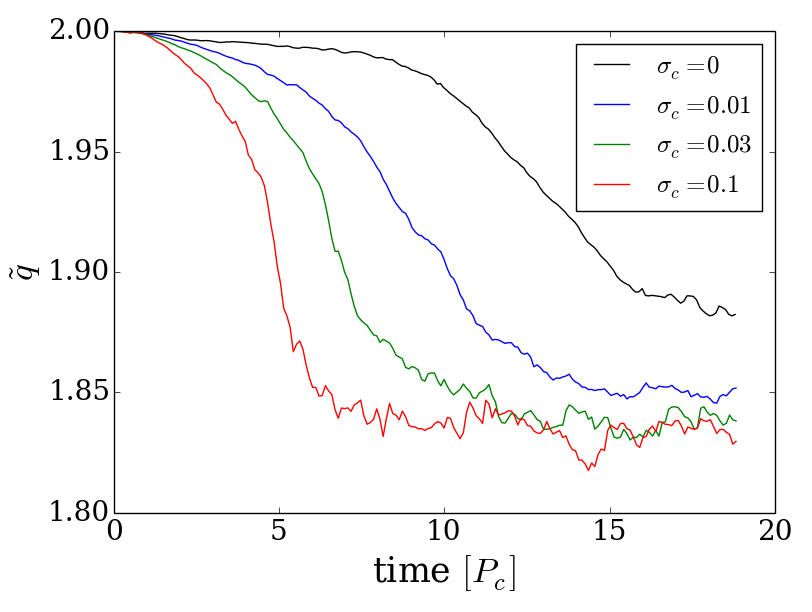}
    \caption{Slope parameter $\tilde{q}$ for models \modela, \modelB, \modelC, and \modelD.}
    \label{fig:q_parameter}
\end{figure}

 All the models we considered produce accretion onto the central black hole, but at different times and in different ways. In the hydrodynamic disc the development of the PPI is the sole responsible for angular momentum transport, hence accretion. After 10 orbital periods, when the $m=1$ mode approaches its maximum amplitude, the kinetic energy and stresses are large (black curve in \refig{fig:energy_stress}), which consequently leads to a significant redistribution of angular momentum (see the evolution of the parameter $\tilde{q}$ in \refig{fig:q_parameter}) and mass loss (almost 30\% of the disc's initial mass). 
 
 The situation is quite different in the magnetized models. The accretion is triggered at much earlier times (after only 2 orbital periods for the model with highest magnetization) with a steeper increase of turbulent kinetic energy and stresses. In model \modelb, the Reynolds and Maxwell stresses increase initially at the same growth rate, but then the magnetic component takes over. For model \modeld~Maxwell stresses dominate from the very beginning and during the whole simulation. Consequently, accretion is enhanced and a higher mass loss results, leading in the most highly magnetized model to a dramatic decrease of the disc mass down to 30\% of its initial value. Concerning the angular momentum distribution, \refig{fig:q_parameter} confirms these findings, showing a faster and earlier decrease of $\tilde{q}$ (see \eqref{eq:tilde_q}) and also a lower saturation value with respect to the hydrodynamic case. 
 
\begin{figure}\centering
    \includegraphics[trim={0 40 0 30},clip,scale=0.3]{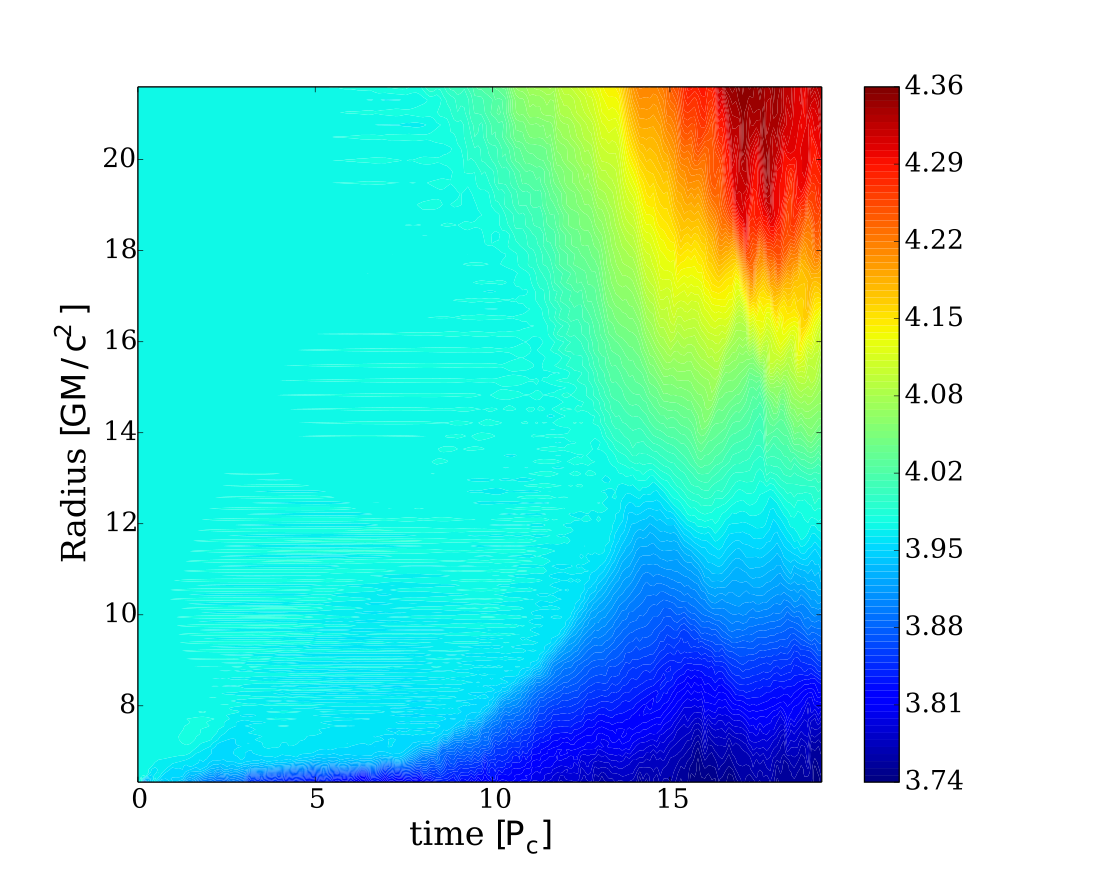}
    \includegraphics[trim={0 40 0 30},clip,scale=0.3]{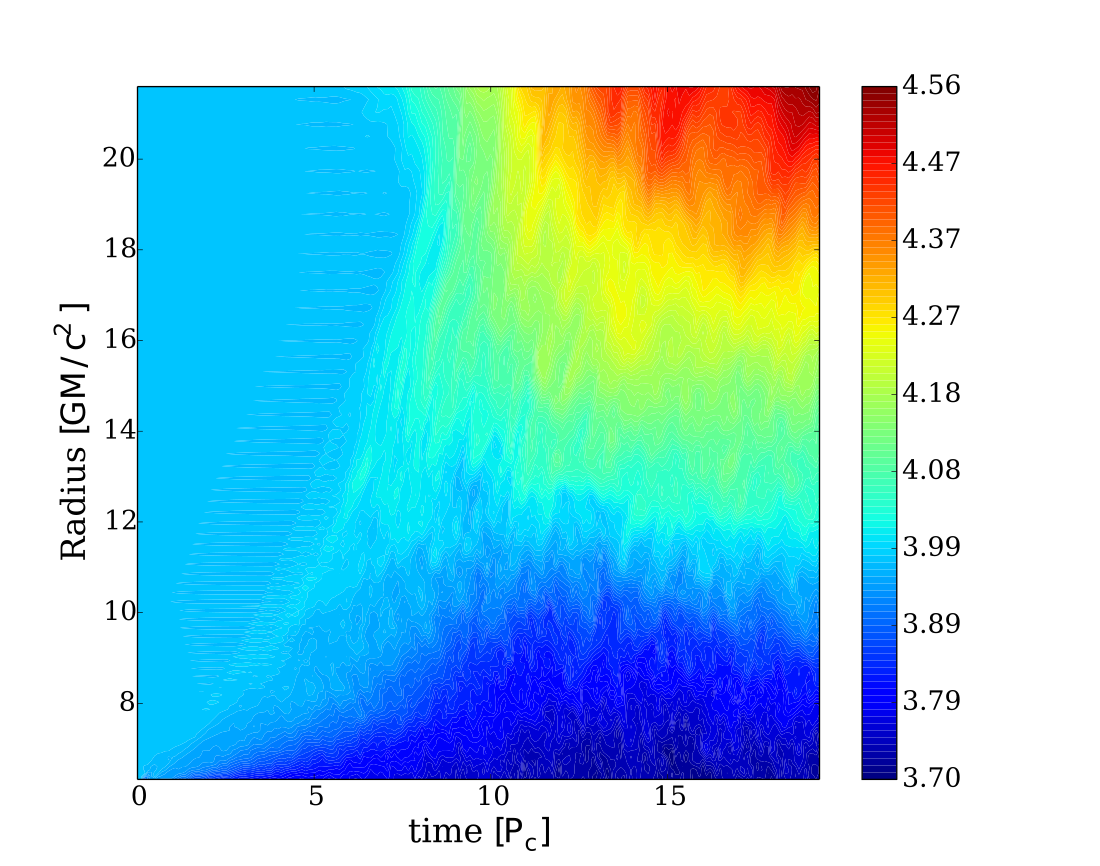}
    \includegraphics[trim={0 10 0 30},clip,scale=0.3]{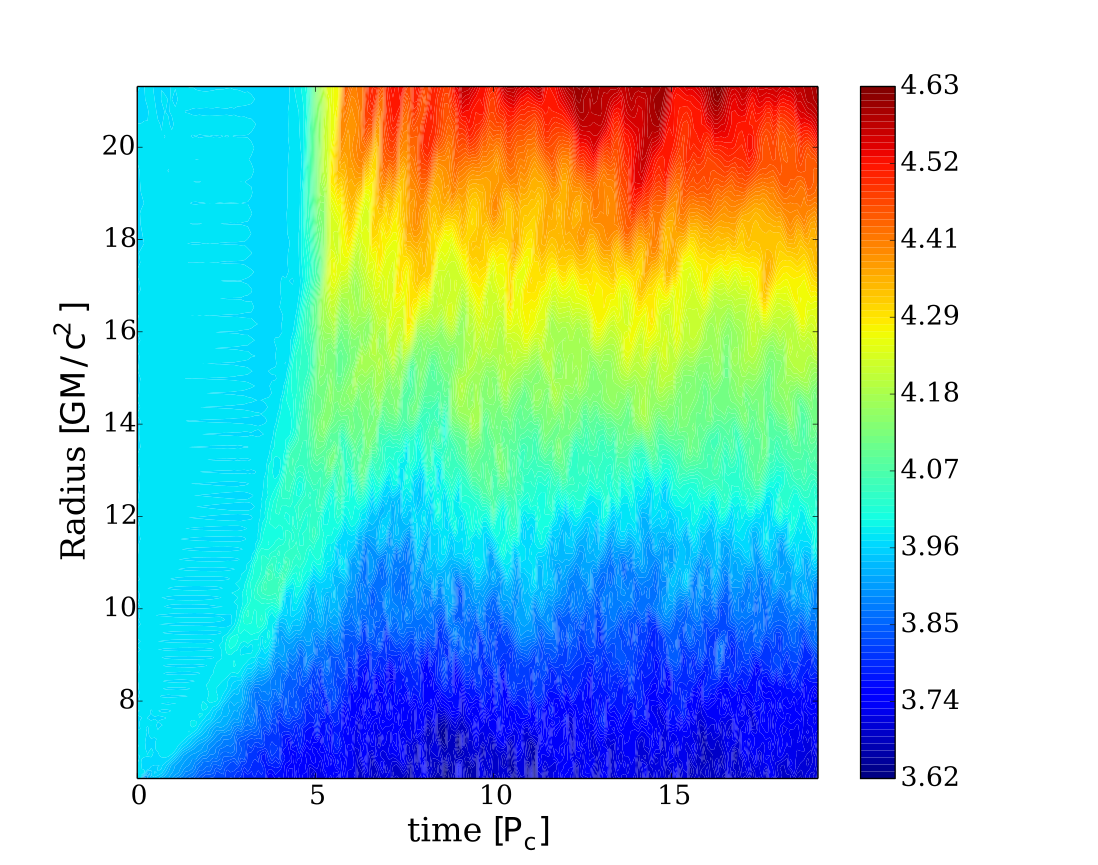}
    \caption{Space-time diagrams of the azimuthally-averaged radial distribution of specific angular momentum $l$ in the equatorial plane for models \modela~(top), \modelB~(middle), and \modelD~(bottom).}
    \label{fig:ang_mom}
\end{figure}

 A space-time diagram of the radial profile of the specific angular momentum $l$ at the equator (top panel in \refig{fig:ang_mom}) clearly shows that in the hydrodynamic model the waves that constitute the unstable mode are increasing $l$ outside the corotation radius and decreasing it in the inner region of the disc, in a non-local fashion.  The same diagrams for models \modelB~and \modelD~(middle and bottom panel of \refig{fig:ang_mom}) show how in the magnetized case angular momentum is transported outwards from the inner regions of the disc much faster than in the absence of a magnetic field. Moreover, there is no substantial trace of a global deposition of angular momentum starting from the outer region of the disc and proceeding inwardly.
  
 \subsection{Mode frequency}

 The mode selected by the PPI as the fastest growing one is characterized not just by the azimuthal number $m=1$, but also by a specific angular frequency $\omega$, which we can measure and directly compare with the disc rotation rate $\Omega$. In fact, the mechanism behind the PPI growth relies on the interaction between a \emph{negative energy} wave (which has $\omega<\Omega$ and therefore is slower than the surroundings) and a \emph{positive energy} wave (with $\omega>\Omega$). This interaction happens at the \emph{corotation radius} $R_{\tx{cor}}$ (defined by $\Omega(R_{\tx{cor}})=\omega$), so that negative energy waves from the inner region of the disc ($r<R_{\tx{cor}}$) exchange energy and angular momentum with the positive energy ones from the outer parts ($r>R_{\tx{cor}}$). A fundamental condition for this process to be efficient is the presence of a reflective boundary \citep{Goldreich:1986} that allows the waves to be reflected and approach the corotation radius where the interaction can take place, leading to a positive feedback and hence the development of the PPI.
 
\begin{figure}
    \includegraphics[trim={1080 0 0 0},clip,scale=0.4]{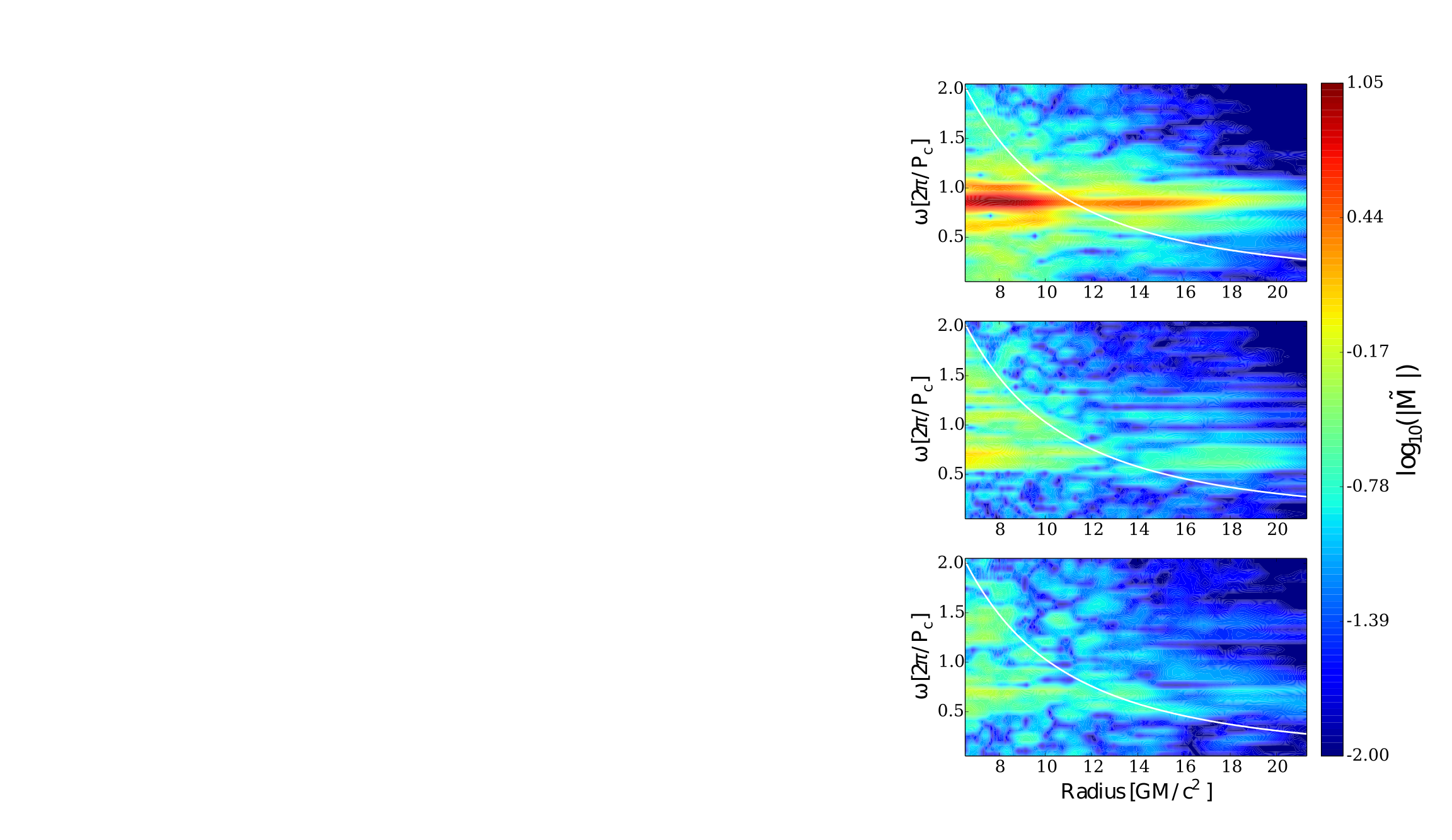}
    \caption{Amplitude of the $m=1$ mode frequency components in a $\omega$ vs. radius diagram calculated from the rest mass density $\rho$. The white curve represents the azimuthally and time averaged equatorial orbital frequency profile. The panels refer respectively to model \modela~(top), \modelb~(middle), and \modelB~(bottom).}
    \label{fig:corotation}
\end{figure}

 The top panel of \refig{fig:corotation} clearly shows the presence of a specific spectral component at $\omega_0\simeq 0.86\Omega_c$ (where $\Omega_c=2\pi/P_c$ is the orbital angular frequency at the disc's center) in the hydrodynamic model, which represents the $m=1$ mode selected by the PPI. Both positive and negative energy waves are present, and they interact through the corotation radius $R_\tx{cor}\simeq r_c$, defined as the location where the mode frequency and the disc's orbital frequency are equal (white curve in \refig{fig:corotation}). The two waves cannot propagate through a narrow forbidden region surrounding the corotation radius, but they can still be transmitted by tunnelling.

 In the magnetized cases (middle and bottom panels) there is no clear selection of a single mode at a well-defined frequency. In the region beyond the corotation frequency the positive-energy waves are much less excited, and there seems to be a lack of modes with frequency below $\omega\sim 0.5\ \Omega_c$.

\subsection{Dependence on resolution}
We focus now on how the results we considered so far change when a lower resolution in azimuthal direction is employed in magnetized models (256 instead of 512 grid points). 

As we can see from the last column of \reftab{tab:hd-models}, some magnetized models have an average value of the quality metric $Q_{\phi}\lesssim 20$, suggesting that in these cases the MHD turbulence might not be sufficiently well resolved. Moreover, considering that the resolution constraint gets stricter when weaker magnetic field are employed, one could also expect the lowly magnetized discs to exhibit greater differences between the two different grids (with respect to the models with higher magnetization).

This expectation is confirmed by \refig{fig:spectra}. The rest mass density of model \modelb~(dashed blue curve in the top panel) shows a small excess of power in the $m=1$ mode, as the ratio $\mathcal{P}_{1,\rho}/\mathcal{P}_{2,\rho}\sim 3$ is roughly twice as large as in the corresponding high-resolution run. Moreover, at all $m$, the power is systematically lower than in the model with the finer grid, pointing to a less efficient MHD-driven mode excitation. This is corroborated by the bottom panel of \refig{fig:spectra}, where the power in the spectrum of the azimuthal Alfv\'en velocity decreases by almost a factor of 2 on the coarser grid for all values of $m<10$.  
The middle panel of \refig{fig:corotation} shows that at lower resolution a still significant excitation of positive-energy modes is present at frequency $\omega\sim 0.7\ \Omega_c$, while there is almost none when a finer grid is employed (bottom panel). The model with low magnetization and $N_{\phi}=256$ displays also a later onset of accretion and growth of stresses, consistent with a less efficient MHD turbulence.

Both models with higher magnetization ($\sigma_c=0.03,0.1$) show no such a significant dependence on the azimuthal resolution. The main difference visible from the averaged spectra in \refig{fig:spectra} is in the slopes at $m\gtrsim 10$, which can be ascribed to the increase in numerical dissipation on the coarser grid for a given accuracy of the reconstruction scheme.   

Both the initial value of $Q_{\phi}$ and the spectral behaviour of the magnetized models suggest that the fastest growing mode of the MRI is underresolved in the low-resolution, low-magnetization model \modelb. In these models the MHD turbulence is not well developed, and the PPI appears to be still capable of exciting to some extent the large-scale $m=1$ mode.   

Since the non-axisymmetric fastest growing modes selected by the MRI are typically associated to short vertical and radial wavelengths \citep{Balbus:1998}, we also expect a dependence of our results on the resolution along the $\theta$-direction. Indeed, running the three magnetized models with a polar resolution decreased by a factor of two leads to results qualitatively similar to those obtained at lower azimuthal resolution. Underresolving the MHD turbulence produces a small excess of power in the low order modes and a general drop in power for modes with $m\gtrsim5$.

\subsection{Dependence on the initial perturbation}
\begin{figure}
    \includegraphics[width=0.49\textwidth]{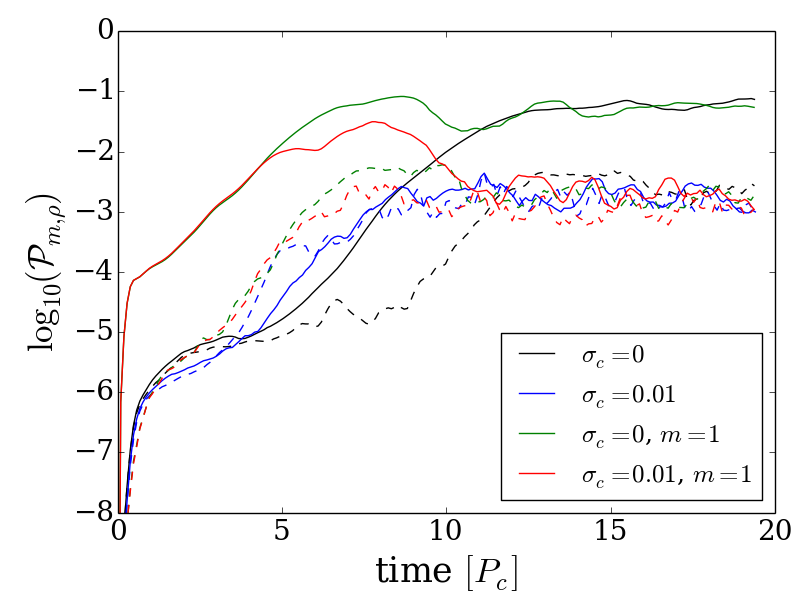}
    \caption{Time evolution of the power in density for the $m=1$ (solid curves) and $m=2$ (dashed curves) modes for models \modela~(black), \modelB~(blue), \modelA~(green), and \modelBm~(red). The former two models are initialized with a random perturbation, while the others are initialized with an $m=1$ perturbation.}
    \label{fig:power_rho_m1}
\end{figure}
So far we considered the evolution of models that were started with a random perturbation of the azimuthal velocity, therefore exciting all modes with a low enough azimuthal number that could be resolved by the number of points in our numerical grid. Our high resolution models are able to resolve the MRI fastest growing mode, which has azimuthal number typically much larger than 1, i.e.
\be
m_\tx{MRI}=\frac{2\pi R}{\lambda_\tx{MRI}}\simeq\frac{R\Omega}{u_\tx{A}^\phi}
\ee
has an average value of $\langle m\rangle_{V}\sim36$ for the models with magnetization $\sigma_c=0.01$. Hence, a random perturbation will equally excite the PPI and the MRI fastest growing modes. However, the growth rate of the fastest MRI mode exceeds the typical growth rate of the hydrodynamic PPI, which plays a key role in the results from the previous sections \citep{Hawley:2000}. 

It is not clear, though, whether the PPI could leave a clear signature on the disk structure if the initial perturbation were to favour the PPI fastest growing mode over the higher order ones excited by the MRI. To assess this, we initialized model \modelB~with a purely $m=1$ perturbation, hence providing the PPI with a \emph{head start} over the MRI. From \refig{fig:power_rho_m1} we can see how in this model (red curve) the $m=1$ mode develops during the first four orbital periods almost exactly as in the hydrodynamic counterpart (green curve). However, the mode's power starts then to drop and approaches the values reached in the randomly perturbed magnetized model (blue curve), i.e. it is damped by almost two orders of magnitude. The power in the $m=2$ mode (dashed curves) is affected in a similar way: its growth significantly slows down with respect to the hydrodynamic case after about five orbital periods, and it reaches values which are roughly a factor of two below the unmagnetized model's ones. 

The spectrograms in \refig{fig:spectrogram_rho} show that high-order modes (i.e. with $m>2$) are excited by the MRI in both models \modelB\ and \modelBm\ (second from the top and bottom panels) roughly at the same time $t\sim 4P_c$, which marks also the beginnining of the departure from the hydrodynamic model's behaviour. Model \modelBm\ still shows a distinctive excitation of the $m=1$ mode (contrary to the randomly perturbed case), which however disappears after $t\sim10P_c$. There is no trace of the episodic excitation of the modes with $2<m<7$ that takes place in the unmagnetized disc between $t=6P_c$ and $t=10P_c$, but instead the power is more uniformly distributed across high-order modes, qualitatively resembling the behaviour of model \modelB.           

\begin{figure}
    \includegraphics[width=0.49\textwidth]{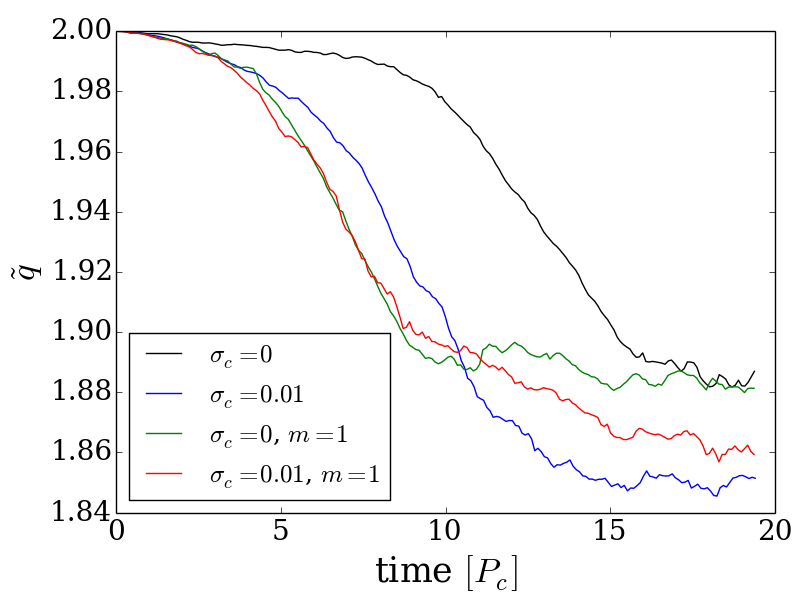}
    \caption{Slope parameter $\tilde{q}$ for models \modela, \modelB, \modelA, and \modelBm.}
    \label{fig:q_m1}
\end{figure} 

\subsection{Interaction between the two instabilities}
From the previous analysis it seems clear that the inclusion of a toroidal magnetic field deeply affects the development of the PPI, even when the field is highly sub-thermal. The $m=1$ mode selected by the PPI is suppressed, leaving no clear signature on the azimuthal or frequency spectra. 

One possible explanation could reside in the early accretion triggered by MHD turbulence, which leads to a faster redistribution of angular momentum across the disc. The free energy stored in the shearing flow decreases and the edge waves that should reach the corotation radius and transport energy are not efficient enough to sustain the unstable $m=1$ mode. \cite{Hawley:2000} found that a torus threaded by a weak constant toroidal magnetic field (i.e. with magnetization $\sigma\sim0.01$) can display an initial appearance of the PPI, though the later development of the MRI interrupts the growth of the hydrodynamic instability. The author suggests the ridistribution of angular momentum produced by the MRI to be the main cause of the inability of the PPI to significantly affect the disc dynamics.

Although a faster transport of angular momentum can play an important role in setting an unfavourable environment for the onset of the PPI, our results suggest that this is not the dominant cause for the lack of large-scale modes when the action of the MRI is taken into account. Indeed, the time evolution of the parameter $\tilde{q}$ for the magnetized model initialized with an $m=1$ perturbation does not dramatically differ from the hydrodynamic counterpart up to $t\sim11P_c$\ (see \refig{fig:q_m1}), although the power in the $m=1$ mode starts showing significant differences already from $t\sim4P_c$. While in the hydrodynamic case $\tilde{q}$ reaches a saturation value of $\sim 1.88$, in model \modelBm~ the transport of angular momentum begins to be mostly dominated by the MRI after ten orbital periods, as $\tilde{q}$ approaches the lower saturation value of $\sim 1.85$ reached by the randomly perturbed magnetized model.   

A comparison between model \modelA\ and \modelBm\ suggests that the main cause of the suppression of large-scale azimuthal modes is actually the presence of high-order modes excited by the MRI, which couple with the $m=1$ mode and put a halt to its growth. The power in the large-scale mode is then effectively damped, leading the disc to a state which is ultimately independent of the initial perturbations. Even when the PPI fastest growing mode is initially enhanced, its later interaction with the small-scale modes caused by the MHD instability ends this transient phase, redistributing power to modes with larger $m$. 

To confirm that in presence of stronger magnetic fields the disc dynamics is completely dominated by the MRI, we performed a run with the exact setup used in model \modeld\ but reduced the azimuthal range to $[0,\pi]$ (model \modele) to filter out the $m=1$ mode from the system. We observed no appreciable change in the evolution with respect to the results obtained using the full azimuthal range $[0,2\pi]$. Hence, we conclude that in this case the dynamics of the system is independent of the large-scale $m=1$ mode, which does not play any significant role in the evolution of our most strongly magnetized model. Powerful accretion and small-scale turbulence set by the MRI prevent any significant development of the PPI fastest growing mode.

These results may seem to be in disagreement with what was reported by \cite{Fu:2011}, who performed an analytical study of the influence of magnetic fields (in both toroidal and poloidal configurations) on the stability on accretion tori. They found that a sufficiently strong toroidal magnetic field can further destabilize the disc and enhance PPI development. However, they assumed incompressibility in their analysis, and therefore they excluded all those modes that instead of being the outcome of the interaction between two surface gravity waves (i.e. the principal branch of the instability) are instead the result of a pressure wave outside the corotation radius coupled to an internal edge wave. Wide tori with an inner edge close to the central black hole can, in fact, be stable to the principal mode. This is due to the fact that the larger is the radial extension of the torus, the slower the external edge wave will be advected by the shear with respect to the corotation radius (because this velocity is proportional to $\Omega$). This delay can be counteracted by the internal edge wave, but only up to a point beyond which the interaction between the two waves can no longer be in phase and the system becomes stable.

Nevertheless, the disc can still be unstable to compressible modes, which are composed of an internal edge-wave and an external pressure wave with radial nodes. Indeed the $m=1$ mode developing in our models does not belong to the principal branch but to the compressible one, where the sound wave composing the mode has two radial nodes. The nature of the mode and the measured growth rate in the hydrodynamic case are consistent with the analysis by \cite{Blaes:1988}. 

\section{Conclusions}
We presented 3D GRMHD simulations of magnetized thick accretion discs around black holes with the goal of estimating the interaction between the PPI, which deeply affects hydrodynamic wide tori with constant specific angular momentum, and the MRI, which is omnipresent in ionized differentially rotating astrophysical discs threaded by magnetic fields. By starting from the magnetized equilibrium solution provided by \cite{Komissarov:2006} we could avoid initial transients in the disc's dynamics and assess how the linear growth and subsequent dynamics of the PPI is affected by the development of the MRI.

Consistently with previous works, we find that even in the presence of a sub-thermal toroidal field the growth of the non-axisymmetric $m=1$ mode (usually the fastest growing mode for wide hydrodynamic tori) is significantly quenched, i.e. no large-scale overdensity structure (planet) forms and the flow is more turbulent. 

The inclusion of toroidal magnetic fields excites higher order modes, thus smaller length-scales, in contrast of the strong dominance of the $m=1$ mode in the purely hydrodynamic case. For all our high-resolution magnetized models there is no clear sign of an excess of power in the $m=1$ mode, showing therefore that its growth has been suppressed by the action of MRI. The redistribution of angular momentum proceeds much faster than in the unmagnetized model, resulting in a lower saturation value for the parameter $\tilde{q}$ measuring the slope of the orbital angular velocity with radius. This is consistent with the fact that, as expected, higher magnetizations lead to stronger stresses and accretion rates onto the black hole.

Launching the model with the weakest magnetic field ($\sigma_c=10^{-2}$) with a monochromatic $m=1$ azimuthal perturbation leads to a transient phase where the PPI fastest growing mode behaves in the same way as in the hydrodynamic countepart. For more than five orbital periods, even in presence of magnetic fields, the disc's dynamics is regulated by the PPI, with the power in the $m=1$ mode that greatly exceeds that of any other azimuthal mode. Then, the power drops and reaches in a very short time the same saturation level as in the randomly perturbed magnetized models. While the ultimate fate of the large-scale mode appears to be the same, independently of the particular spectrum of the perturbation, we observe nevertheless a transient PPI dominated phase that lasts for a few orbital periods and that could be interesting in those astrophysical scenarios where the gravitational interaction with the central black hole can indeed excite the $m=1$ mode.

The comparison between model \modelA and \modelBm offers a very interesting insight into the physical mechanism behind the suppression of the PPI. The redistribution of specific angular momentum from a constant radial profile towards the Keplerian one occurs for both the PPI and the MRI, and it has been invoked in previous studies \citep{Hawley:2000} as the sole responsible of the missing development of the PPI in three-dimensional MHD global simulations, since the hydrodynamic instability is intrinsically susceptible to being stabilized if enough angular momentum is transported outwards in the disc. Our results show, however, that this mechanism is not the main agent of the suppression of the PPI. When quenching of the initially excited $m=1$ mode occurs in model \modelBm, there is no sign of a significant change in the specific angular momentum radial distribution with respect to the hydrodynamic case. Only after the power in the large-scale mode is damped down to the saturation level of the other magnetized models one can notice an incipient significant drop in the parameter $\tilde{q}$. We identify, therefore, the coupling of the $m=1$ mode with higher order modes excited by the MRI as the main mechanism that quenches the growth of the PPI and leads to a turbulent state dominated by the MRI and ultimately independent of the initial perturbation.

The initial PPI dominated evolution of the magnetized model \modelBm~ is consistent with the findings of \cite{Nealon:2017}, who argued that a thick disc can undergo a significant accretion phase dominated by the PPI, before the onset of the MRI. Their analysis is based on numerical measurements of accretion rates in the evolution of the remnant of a hydrodynamic tidal disruption event around a supermassive black hole and a comparison between the expected duration of the linear phases of the PPI and MRI. They argue that for weak enough magnetic fields the saturation time of the MRI will significantly exceed that of the PPI, suggesting that in that case the disc could undergo an accretion phase regulated mostly by the PPI. However, they also notice that using this argument to discern between a PPI and MRI dominated regime can lead to a significant overestimate of the magnetic field strength required for the latter scenario. Indeed, we found that the quenching of the PPI fastest growing mode starts well before the saturation of either of the instabilities.

Our results suggest that with the inclusion of magnetic fields the dominance of the $m=1$ mode (with respect to higher order ones) due to PPI should not hold in thick tori, but there are some caveats that need to be addressed. First, we neglected the self-gravity of the disc. This is expected to be a good approximation for example in X-ray binaries, but not in the case of the remnant of a NS-NS merger that produces a black-hole-torus system. It has been shown by \cite{Korobkin:2011} that the gravitational interaction between the disc and the central black hole can indeed further excite the $m=1$ mode, leading hence possibly to a different outcome than ours once a magnetic field is included into the simulations.

In this work we considered tori with a constant specific angular momentum radial profile, which are most susceptible to PPI and therefore represent the most favourable environment for its development. Despite the general suppression by the MRI, we find an initial transient phase where the PPI dominates the evolution of a magnetized disc excited with a weak $m=1$ perturbation. It would be interesting to assess whether this would still be the case when using magnetized equilibrium solutions with a more realistic distribution of angular momentum, such as those recently presented in \cite{Gimeno-Soler:2017}.

Another important aspect is the role of turbulent resistivity. We found that if the dynamical evolution of the MRI is not resolved properly, the system shows a residual excess of power in the $m=1$ mode and a clear signature of negative- and positive-energy waves coupled through the corotation radius. This suggests that for a strong enough magnetic dissipation, which could be present in a turbulent environment, the MRI could be quenched, and hence the PPI could still grow significantly and produce a non-negligible $m=1$ like overdensity. As mentioned in Section 2, in a forthcoming work we will investigate how the system is affected by an explicit magnetic diffusivity.

\section{Acknowledgments}
The authors would like to thank C.~Gammie, O.~Blaes, O.~Sadowski, M.~Wielgus, R.~Nealon and T.~Font for providing helpful comments and suggestions. A special thank to the anonymous referee, whose comments helped us improving this manuscript.

MB thanks, in particular, F.~Baruffa and M.~Rampp for their invaluable assistance and support during the development of the parallelization scheme used in \texttt{ECHO}. The simulations were carried out on the Hydra cluster at the Max Planck Computing and Data Facility (MPCDF) and on the SuperMuc cluster at the Leibniz-Rechenzentrum (LRZ) of the Bavarian Academy of Sciences.

JG acknowledges support from the Max-Planck-Princeton Center for Plasma Physics and from the European Research Council (grant MagBURST--715368).

\appendix
\section{Coordinate transformation}
To exploit the regularity property of the Kerr-Schild (KS) coordinates at the event horizon, we perform a transformation of all the primitive variables computed for a stationary torus in Boyer-Lindquist (BL) coordinates. In the following, the labels KS and BL indicate quantities measured by a KS and BL \emph{Eulerian observer} respectively, while unprimed and primed indices refer to quantities expressed in KS and BL \emph{coordinates}.\\
One has to consider the linear transformation:

\begin{equation}\label{lin_transf}
 \mathcal{A}^\mu_{\mu'}=\begin{pmatrix}
                         1 & G & 0 & 0 \\
                         0 & 1 & 0 & 0 \\
                         0 & 0 & 1 & 0 \\
                         0 & H & 0 & 1
                        \end{pmatrix}
                        \textrm{ with } \left\{ \begin{matrix}
                                                 G=-\frac{2r}{\Delta} \\
                                                 H=-\frac{a}{\Delta}
                                                \end{matrix}\right.
\end{equation}
where $\Delta=r^2-2Mr+a^2$, $r$ is the radial coordinate, and $a$ is the black hole spin (which in general will be different from 0). This transformation relates vectors and tensors respectively:
\begin{align}
 x^\mu&=\mathcal{A}^\mu_{\mu'} x^{\mu'}, \\
 T^{\mu\nu}&=\mathcal{A}^\mu_{\mu'}\mathcal{A}^\nu_{\nu'}T^{\mu'\nu'}.
\end{align}
We also have to consider that Eulerian observers in the two coordinate systems are not identical in general, i.e. we cannot simply apply the transformation in \eqref{lin_transf} to the vectorial primitive variables $(\bm{v}, \bm{B}, \bm{E})$, which represent quantities measured in the Eulerian frame of reference. For quantities like the fluid 4-velocity $u^\mu$ and the Faraday tensor $F^{\mu\nu}$, on the other hand, one must only apply \eqref{lin_transf} to fully take into account the change of frame of reference (e.g., $\mathcal{A}^\mu_{\mu'}\bl{u^{\mu'}}=\ks{u^\mu}$). 

To obtain for $\bm{v}$, $\bm{E}$, and $\bm{B}$ the correct relation between KS components measured by the KS Eulerian observer and BL components measured by the BL Eulerian observer, we first write the spatial velocity, magnetic field, and electric field in terms of 4-velocity and Faraday tensor in KS coordinates as measured by the KS Eulerian observer:
\begin{equation}
 \ks{v^i}=\frac{1}{\ks{\alpha}}\left(\frac{\ks{u}^i}{\ks{u}^t}+\ks{\beta}^i\right),
\end{equation}
\begin{equation}
 \ks{B^i}=\ks{\alpha}\ks{F^{*ti}},
\end{equation}
\begin{equation}
 \ks{E^i}=\ks{\alpha}\ks{F^{ti}}.
\end{equation}
Next we use \eqref{lin_transf} to transform the components of $\ks{u^i}$, $\ks{F^{*ti}}$ and $\ks{F^{ti}}$ in BL coordinates, then we write these components in terms of $\bl{v}^{i'}$, $\bl{E^{i'}}$ and $\bl{B^{i'}}$. The result gives the correct transformation rules:
\begin{equation}
\ks{v^i}=\frac{1}{\ks{\alpha}}\left[ \frac{\mathcal{A}^{i}_{j'}(\bl{\alpha}\bl{v}^{j'}-\bl{\beta}^{j'})}{1-\mathcal{A}^0_{r'}\bl{\alpha}\bl{v}^{r'}}+\ks{\beta}^i\right], 
\end{equation}
\begin{equation}
\ks{B^i}=\frac{\ks{\alpha}}{\bl{\alpha}} \left\{ \bl{B}^i-\mathcal{A}^0_{r'}\left[ \bl{B}^{r'}\bl{\beta}^i+\frac{\bl{\alpha}}{\bl{\gamma}^{1/2}}\left(\delta^i_{\phi'}E^{\rm\mbox{\tiny BL}}_{\theta'}-\delta^i_{\theta'}E^{\rm\mbox{\tiny BL}}_{\phi'}\right)\right] \right\},
\end{equation}
\begin{equation}
\ks{E^i}=\frac{\ks{\alpha}}{\bl{\alpha}} \left\{ \bl{E}^i-\mathcal{A}^0_{r'}\left[ \bl{E}^{r'}\bl{\beta}^i-\frac{\bl{\alpha}}{\bl{\gamma}^{^1/2}}\left(\delta^i_{\phi'}B^{\rm\mbox{\tiny BL}}_{\theta'}-\delta^i_{\theta'}B^{\rm\mbox{\tiny BL}}_{\phi'}\right)\right] \right\}.
\end{equation}

\section{Primitive variables recovery stability}
The most delicate and at the same time computationally expensive part of the numerical algorithm usually implemented in a standard GRMHD code is the inversion from conservative to primitive variables. During this procedure it is possible for the numerical algorithm to fail in providing physically acceptable values for the gas pressure, i.e. non-negative ones, whenever the magnetization value in the grid cell is so high that the magnetic energy density $U_{\tx{em}}=\tfrac{1}{2}(E^2+B^2)$ almost matches the total energy density $U$. In these cases the numerical evaluation of the quantity $U-U_{\tx{em}}$ could produce a negative number, and hence a negative pressure can be encountered. To avoid this, \texttt{ECHO} evolves along with the standard GRMHD equations also the conservation law for the entropy density:
\be\label{eq:adiab}
\nabla_\mu(\rho su^\mu)=0
\ee
where $s=p/\rho^\Upsilon$. The adiabatic condition \eqref{eq:adiab} is equivalent to the total energy conservation in absence of shocks or other sources of dissipation, and hence it is used whenever the use of the energy equation fails. We found that this procedure solved the vast majority of numerical issues related to this part of the algorithm. In the very few cases when this solution does not work either, we simply reset the primitives to their values at the previous time step.

\bibliographystyle{mn2e}
\bibliography{./bugli17}

\end{document}